\documentclass[aps,amsmath,amssymb,showpacs,showpacs,superscriptaddress,pre]{revtex4-1}

\usepackage{graphicx,color}
\usepackage{bm}
 \graphicspath{{./}{./figures/}}

\begin{document}
\title{Survival 
Probability of Random Walks and L\'evy Flights on a Semi-Infinite Line}

\author{Satya N. Majumdar}
\email[]{majumdar@lptms.u-psud.fr}
\affiliation{LPTMS, CNRS, Univ. Paris-Sud, Universit\'{e} Paris-Saclay, 91405 Orsay, France}
\author{Philippe Mounaix}
\email[]{philippe.mounaix@cpht.polytechnique.fr}
\affiliation{Centre de Physique Th\'eorique, Ecole
Polytechnique, CNRS, Universit\'e Paris-Saclay, F-91128 Palaiseau, France}

\author{Gr\'egory Schehr}
\email[]{gregory.schehr@lptms.u-psud.fr}
\affiliation{LPTMS, CNRS, Univ. Paris-Sud, Universit\'{e} Paris-Saclay, 91405 Orsay, France}
\begin{abstract}
We consider a one-dimensional random walk (RW) with a continuous and symmetric jump distribution, $f(\eta)$, characterized by a L\'evy index $\mu \in (0,2]$, which includes standard random walks ($\mu=2$) and L\'evy flights ($0<\mu<2$). We study the survival probability, $q(x_0,n)$, representing the probability that the RW stays non-negative up to step $n$, starting initially at $x_0 \geq 0$. Our main focus is on the $x_0$-dependence of $q(x_0,n)$ for large $n$. 
We show that $q(x_0,n)$ displays two distinct regimes as $x_0$ varies: (i) for $x_0= O(1)$ (`quantum' regime), the discreteness of the jump process significantly alters the standard scaling behavior of $q(x_0,n)$ and (ii) for $x_0 = O(n^{1/\mu})$ (`classical' regime) the discrete-time nature of the process is irrelevant and one recovers the standard scaling behavior (for $\mu =2$ this corresponds to the standard Brownian scaling limit). The purpose of this paper is to study how precisely the crossover in $q(x_0,n)$ occurs between the quantum and the classical regime as one increases $x_0$. 
\end{abstract}

\maketitle

\section{Introduction}

\begin{figure}[ht]
\includegraphics[width = 0.7\linewidth]{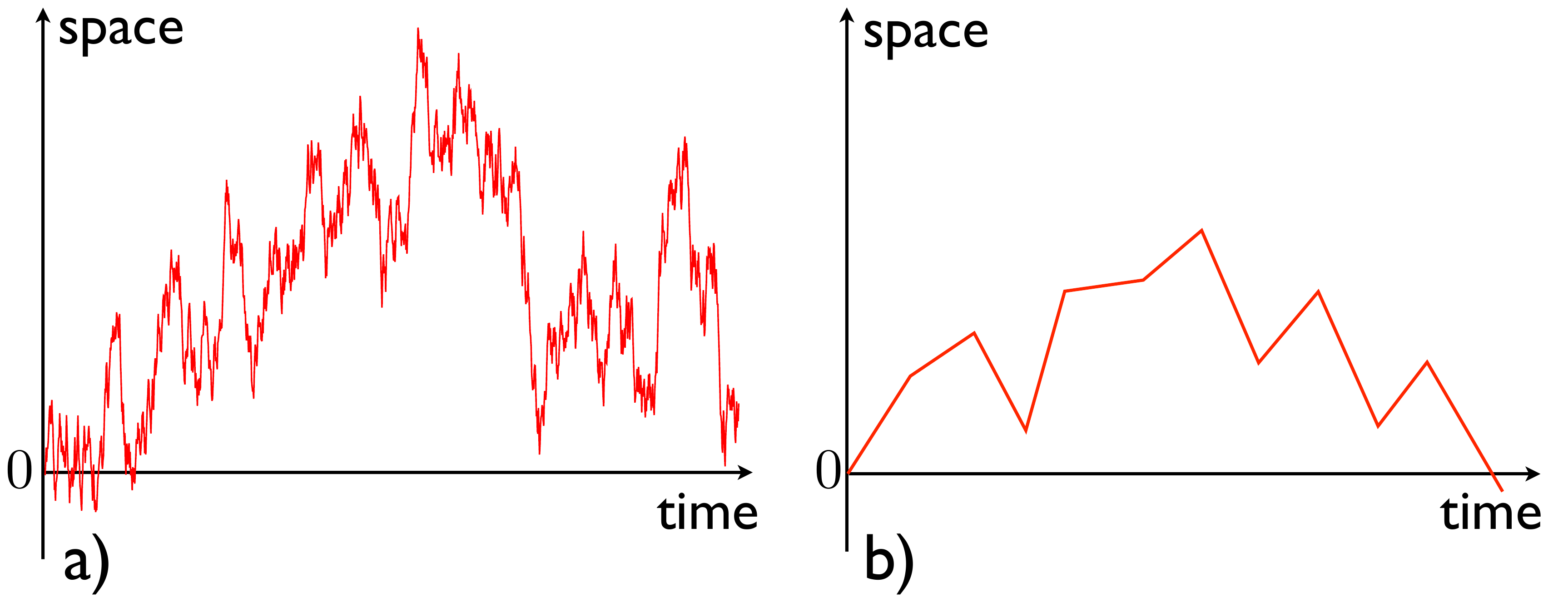}
\caption{{\bf a):} Typical trajectory of a Brownian motion (\ref{brown.1}) starting from 
the origin $x_0 = 0$. It immediately crosses and re-crosses the origin infinitely often, yielding a vanishing survival probability $q(x_0,t)$, as in Eq. (\ref{qxt_asymp}). {\bf b)} Typical trajectory of a discrete-time random walk starting from the origin $x_0=0$ and staying positive right after. Since the walker can travel for several steps before first crossing the origin, the survival probability $q(0,n)$ is finite, as in Eq. (\ref{SA.3}).}\label{Fig_intro}
\end{figure}

Consider a simple Brownian walker on a line whose position $x(t)$ evolves, starting initially at $x_0>0$, in continuous time
via the Langevin equation
\begin{equation}
\frac{dx}{dt}= \eta(t)
\label{brown.1}
\end{equation}
where $\eta(t)$ is a Gaussian white noise with zero mean and a correlator
$\langle \eta(t)\eta(t')\rangle= 2\, D\, \delta(t-t')$. Let $q(x_0,t)$ 
denote
the probability that the walker does not cross zero up to time $t$, starting
at $x_0>0$ at $t=0$. This is called the persistence or the survival 
probability
of the walker and has been extensively studied in the 
literature~\cite{Feller,Redner_book,Satya_review,Bray_review,AS_review}.
In fact, $q(x_0,t)$ satisfies a backward Fokker-Planck equation where one considers $x_0$ as a 
variable~\cite{Satya_review,Bray_review}
\begin{equation}
\frac{\partial q(x_0,t)}{\partial t}= D\, \frac{\partial^2 q(x_0,t)}{\partial x_0^2}\,
\label{bfp.1}
\end{equation}
valid for $x_0\ge 0$ with the absorbing boundary condition $q(x_0=0,t)=0$ at the origin and with
the initial condition $q(x_0>0,t=0)=1$. The solution is simply
\begin{equation}
q(x_0,t)= {\rm erf}\left(\frac{x_0}{\sqrt{4\,D\,t}}\right); 
\quad {\rm where} \quad {\rm erf}(z)= \frac{2}{\sqrt{\pi}}\, \int_0^z e^{-u^2}\, du\,.
\label{erf.1}
\end{equation}
Using ${\rm erf}(z)\approx 2z/\sqrt{\pi}$ as $z\to 0$, it follows that for any fixed $x_0\ge 0$, the
survival probability at late times decays as a power law
\begin{equation}
q(x_0,t) \sim \frac{x_0}{\sqrt{\pi Dt}}\ \ \ \ \ {\rm for}\ t\rightarrow +\infty .
\label{qxt_asymp}
\end{equation}
Thus, for any fixed $x_0\ge 0$, the walker eventually crosses the origin when the
time $t$ exceeds the characteristic diffusion time $t^*= x_0^2/2D$. In particular, if the walker
starts at the origin $x_0=0$, $t^*=0$ and the walker dies immediately with probability $1$.
In other words, $q(x_0=0,t)=0$ at all times. This shows up in any typical Brownian trajectory starting at 
the origin at $t=0$. It immediately crosses and re-crosses the origin infinitely often, making it
impossible for the walker to survive (see Fig. \ref{Fig_intro} a) for a typical Brownian trajectory).

Consider now a random walker on a line that evolves in discrete time by making random independent jumps
at each time step. Starting from the initial position 
$x_0$, the position of the walker now evolves in 
discrete time via the simple Markov jump rule
\begin{equation}
x_n= x_{n-1} +\eta_n
\label{markov.1}
\end{equation}
where $\eta_n$ represents the random jump at step $n$. The jump lengths
$\eta_n$'s are assumed to be independent and identically distributed 
(i.i.d.) random variables, each drawn from a continuous and symmetric
probability distribution function (PDF), $f(\eta)$, the Fourier transform of which
\begin{equation}
{\hat f}(k)= \int_{-\infty}^{\infty} f(\eta)\, e^{ik\eta}\, d\eta
\label{ft.1}
\end{equation}
is assumed to have the following small-$k$ behavior
\begin{equation}
{\hat f}(k)= 1- (a_\mu\,|k|)^{\mu}+\ldots
\label{smallk.1}
\end{equation}
where $0< \mu\le 2$ and $a_\mu$ represents a typical length scale 
associated
with the jump.
The L\'evy exponent $0<\mu\le 2$ dictates
the large $|\eta|$ tail of $f(\eta)$. For jump PDF's with a finite
second moment $\sigma^2= \int_{-\infty}^{\infty} \eta^2\, f(\eta)\,d\eta$,
such as Gaussian, exponential, uniform etc,
one evidently has $\mu=2$ and $a_2=\sigma/\sqrt{2}$.
In contrast, $0<\mu<2$
corresponds to jump densities with fat tails
$f(\eta)\sim |\eta|^{-1-\mu}$ as $|\eta|\to \infty$, generally 
called L\'evy flights with index $\mu$ (for reviews on these jump 
processes see~\cite{BG90,MK00}). 

Let $q(x_0,n)$ now denote the persistence, or survival probability, of this
discrete-time walker starting at $x_0\ge 0$ up to time $n$. Unlike in the continuous-time 
Brownian motion and as a consequence of the discrete-time dynamics of the jump process, there is now a finite fraction of trajectories starting at $x_0=0$ that travel several steps before first crossing the origin to the negative side [see Fig. \ref{Fig_intro} b)], yielding a non-zero $q(x_0=0, n)$ at any finite $n$. For a continuous and symmetric $f(\eta)$, $q(0,n)$ is given by the universal
Sparre Andersen formula~\cite{SA}
\begin{equation}
q(0,n)= {2n\choose n}\, 2^{-2n}\, \,
\label{SA.2}
\end{equation}
which decays algebraically for large times $n$ as  
\begin{equation}
q(0,n) \sim \frac{1}{\sqrt {\pi\, n}}\ \ \ \ \ {\rm for}\ n\rightarrow +\infty.
\label{SA.3}
\end{equation}
Let us emphasize that the results in Eqs. (\ref{SA.2}) and (\ref{SA.3}) are completely universal, i.e,
independent of $f(\eta)$ and the $1/\sqrt{n}$ algebraic decay holds for all L\'evy flights with index $0<\mu\le 2$.

The comparison between Eqs.\ (\ref{qxt_asymp}) and\ (\ref{SA.3}) raises some questions. Consider, for instance, a discrete-time walk with jumps of finite variance $\sigma^2$, i.e. with index $\mu=2$ and $a_2=\sigma/\sqrt{2}$ in Eq. (\ref{smallk.1}).
For such a walk, central limit theorem tells us that the discrete-time process $x(n)$ converges for large $n$ to
the continuous-time Brownian motion $x(t)$, upon identifying $2\, D \,t= \sigma^2 n$. Hence, the persistence $q(x_0,n)$
should converge to the Brownian motion result. As a result, one may naively replace $t$ by $n\sigma^2 /2D$ in
the Brownian result in Eq. (\ref{qxt_asymp}) and conclude that the $1/\sqrt{n}$ decay given by the Sparre Andersen
theorem in Eq. (\ref{SA.3}) is basically the same as the $1/\sqrt{t}$ decay of persistence for the Brownian motion. There
are however two problems with this simplistic
picture: (i) the $1/\sqrt{n}$ behavior in Eq.\ (\ref{SA.3}) holds not just for Brownian motion,
but also for L\'evy flights with divergent $\sigma^2$ (i.e., $\mu<2$) which do not converge to Brownian motion at late times.
Hence the $1/\sqrt{n}$ decay in Sparre Andersen theorem has a different origin than the $1/\sqrt{t}$ decay of the Brownian persistence.
(ii) More importantly, the Brownian persistence vanishes when $x_0\to 0$ in Eq. (\ref{qxt_asymp}), i.e., $q(x_0=0,t)=0$, while
the persistence $q(x_0,n)$ for the discrete-time jump process remains finite even when $x_0=0$ (as in Eq. (\ref{SA.3})).
So, how would it be possible to reconcile between these seemingly different results for persistence in the discrete and continuous time processes?    

The resolution to this puzzle is actually simple. There are two ways to take the continuous time limit of the discrete-time persistence 
$q(x_0,n)$ for jump PDF's with finite variance $\sigma^2$. Either one fixes $x_0$ and takes the limits $n\to \infty$ and $\sigma^2\to 0$ keeping the product $\sigma^2\, n= 2\,D\,t$ fixed, where $t$ is the continuous time, or one fixes $\sigma^2$ and scales $x_0\sim \sqrt{n}$ for large
$n$ in order to converge to the Brownian result. Essentially, what matters is that the ratio
$x_0/\sqrt{\sigma^2\, n}$ should be kept $O(1)$. Let us work with fixed $\sigma^2$. Then, the Brownian result in Eq. (\ref{erf.1})
is recovered only in the scaling limit when $x_0\sim \sqrt{n}$, i.e., for large $x_0$. For $x_0= O(1)$, there is no
convergence to the Brownian limit, hence the universal result $q(x_0, n)\sim 1/\sqrt{\pi\, n}$ as $x_0\to 0$ 
(Sparre Andersen limit) has
nothing to do with the $1/\sqrt{t}$ Brownian decay which holds only when $x_0\sim \sqrt{n}$.

One is then faced with the following question: how precisely does the large $n$ behavior of $q(x_0,n)$ as a function of $x_0$ (for jump processes with a finite $\sigma^2$) cross over 
from the Sparre Andersen limit ($x_0\to 0$) to the Brownian limit ($x_0\sim \sqrt{n}$) as $x_0$ is increased? 
The main purpose of this paper is to address this question.
We will show that there are indeed two different scales of $x_0$, namely $x_0= O(1)$ and $x_0\sim n^{1/2}$, where
the large $n$ scaling behaviors of $q(x_0,n)$ are very different. We 
will call the first regime (with $x_0= O(1)$) 
the {\it discrete `quantum'}
regime, as the discrete-time nature of the process plays a dominant role in this regime. In contrast the second regime
(with $x_0\sim n^{1/2}$) will be referred to as the {\it 'classical'} scaling regime, i.e., the usual Brownian scaling regime. A similar picture of two
separated scales of $x_0$ appears for jump processes with a divergent $\sigma^2$ (i.e., L\'evy flights with $0<\mu<2$). Here
again the large $n$ behavior of $q(x_0,n)$ is different in the two regimes: $x_0= O(1)$ (discrete `quantum' regime), 
and $x_0 \sim n^{1/\mu}$ (`classical' scaling regime). This latter regime  reduces to the Brownian scaling regime if $\mu=2$. 
In this paper, we will study $q(x_0,n)$ for general $0<\mu \le 2$. 
We will compute the large $n$ behavior of $q(x_0,n)$ in both `quantum' and standard scaling regimes and we will demonstrate how
the rather different results in these two regimes match smoothly as $x_0$ is increased from $O(1)$ to $O(n^{1/\mu})$.

This clarification of the asymptotic behavior of $q(x_0,n)$ in two widely separated scales of $x_0$ 
is particularly important in the light of several
current applications of random walks, where 
the persistence
probability $q(x_0,n)$ turns out to be a crucial ingredient or a building 
block. {{ In fact, the earliest application of this half-space problem
in presence of an absorbing boundary goes back to the celebrated `Milne' problem in 
astrophysics in connection with the
scattering of light from the sun's surface~\cite{Milne}, and later a similar
problem appeared in
the theory of transport of neutrons through a non-capturing medium~\cite{PS47,finch}. 
In chemistry, the survival probability $q(x_0,n)$ is an important
observable in the study of diffusion in presence of an absorbing 
boundary \cite{ZK95}. 
Amongst more recent applications, the knowledge of the
asymptotic behavior of $q(x_0,n)$ was found to be crucial in computing
the precise 
statistics 
of the global 
maximum of a random walk evolving via Eq.~(\ref{markov.1})~\cite{CM2005} }}. 
Indeed, if one denotes by $x_{\max}$ the global maximum of a random walk  
evolving via Eq.~(\ref{markov.1}) starting from the origin, the survival 
probability $q(x_0,n)$ actually coincides with the cumulative distribution of the maximum 
$x_{\max}$, 
i.e., ${\rm Prob.}(x_{\max}\leq x_0) = q(x_0,n)$~\cite{CM2005,M2010}. The 
same quantity $q(x_0,n)$ was 
also shown to
appear in the problem of the capture of particles into a spherical
trap in $3$-dimension, known as the Smoluchowski 
problem~\cite{MCZ2006,ZMC2007}. It also plays an important role in computing the order and gap statistics
of a random walk sequence, e.g., in calculating the distribution
of the gap between the $k$-th and the $(k+1)$-th maxima of a random
walk sequence in Eq. (\ref{markov.1}) (see Refs.~\cite{SM2012,SM_review}). Similarly,
the joint distribution of the gap and time-lag between the highest and the 
second highest maximum, both for a discrete-time random walk
sequence in Eq. (\ref{markov.1}), as well as for 
the so called continuous-time 
random walk (CTRW), requires a precise knowledge of 
$q(x_0,n)$~\cite{MMS2013,MMS2014,MSM2016,MS2017}.
Finally, $q(x_0,n)$ is at the heart of fluctuation theory \cite{Feller,Bin2001} and plays
a major role in the statistics of records and associated observables in 
several correlated time-sequences
generated from the basic simple random walk sequence in Eq. 
(\ref{markov.1})~\cite{Feller,Bin2001,MZ2008,WMS2012,MSW2012,GMS2014,GMS2015,GMS2016}
(for a recent review on record statistics, see Ref. \cite{record_review}).
Hence clarifying the precise asymptotic properties of $q(x_0,n)$
in different regimes of $x_0$ is important and crucial.

Let us remark that many results on the asymptotic properties of 
$q(x_0,n)$ for large $n$ are already known, in particular in the limit $x_0=0$ 
(see e.g. Ref.~\cite{SA,Feller,Redner_book,Bray_review}) and in the scaling regime when 
$x_0\sim n^{1/\mu}$ for large $n$ (see for instance, 
Refs.~\cite{CM2005,MCZ2006,M2010,WMS2012}). However, how these two 
asymptotic behaviors match precisely as $x_0$ increases from $O(1)$ to 
$O(n^{1/\mu})$ has not been clearly elucidated yet, to the best of our 
knowledge, for general jump processes with $0<\mu\le 2$. 
This issue was briefly addressed in Ref.~\cite{MCZ2006} in a somewhat
different context (see also Ref.~\cite{M2010} for a discussion),
but only for the special case of
exponential jump distribution, i.e., $f(\eta)= (1/2)\, e^{-|\eta|}$.
This paper gathers in one place the scattered 
literature on $q(x_0,n)$ for a L\'evy flight, with arbitrary $0<\mu\le 2$, 
on a semi-infinite
line in the presence of an absorbing boundary at the origin
and present a
unifying picture 
that demonstrates the matching of  
$q(x_0,n)$ across the two widely separated scales of $x_0$, i.e. 
$x_0= O(1)$ and $x_0= O(n^{1/\mu})$. {{ While the spirit of 
this manuscript is thus more of a review, there are nevertheless some new results
as well, a list of which can be found in the concluding section.}}

The rest of the paper is organized as follows. In Section II, we 
summarize our main results. 
In Section III, we provide the general setting for calculating
the survival probability using the Pollaczek-Spitzer formula.
In Section IV, we discuss the discrete quantum regime when $x_0= 
O(1)$. In Section V, we consider the classical scaling regime.
Finally, we conclude with a discussion and {{a list of the new results}} in Section 
VI. {{ Some technical details of the computations are relegated to the Appendices A and B.}}

\section{Summary of Main Results}

\begin{figure}
\includegraphics[width = \linewidth]{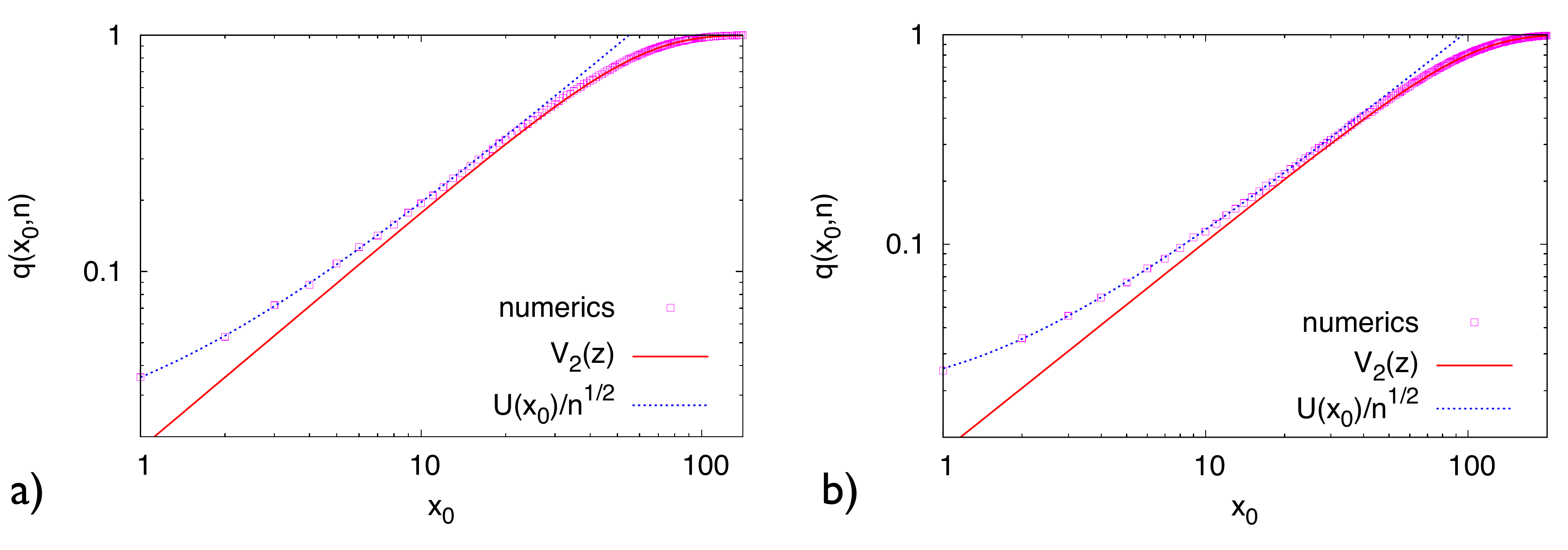}
\caption{{{\bf a)} Log-log plot of $q(x_0,n)$ for the exponential jump distribution $f(\eta) = (1/2)e^{-|\eta|}$ and $n = 1000$ steps. The squares represent numerical results. The solid red line corresponds to $V_2(z = x_0/\sqrt{n})$ given in Eq. (\ref{V2z.1}) with $a_2 = 1$ while the dotted blue line corresponds to $U(x_0)/\sqrt{n}$ with $U(x_0)$ given in Eq. (\ref{ux_exp}) with $a=1$. {\bf b)} The same quantity as in the left panel, $q(x_0,n)$, plotted on a log-log plot for a gamma jump distribution $f(\eta) = (1/2)|\eta| \, e^{-|\eta|}$ and $n = 1000$ steps. The solid red line corresponds to $V_2(z = x_0/\sqrt{n})$ given in Eq.~(\ref{V2z.1}) with $a_2 = \sqrt{3}$ while the dotted blue line corresponds to $U(x_0)/\sqrt{n}$ with $U(x_0)$ given in Eq.~(\ref{ux_xexp}) with $a=\sqrt{3}$.
These plots clearly illustrate the two different regimes for $x_0 = O(1)$ and $x_0 = O(\sqrt{n})$ as well as the crossover from the former to the latter as $x_0$ increases past $x_0\sim 30$}.}\label{Fig_crossover}
\end{figure}

In this Section we summarize our main results.  
We consider the large $n$ 
behavior
of the persistence $q(x_0,n)$ for arbitrary, {{ {\it symmetric and continuous} }} jump 
PDF $f(\eta)$ whose
Fourier transform ${\hat f}(k)$ behaves for small-$k$ as in Eq. 
(\ref{smallk.1}), with L\'evy index $0<\mu\le 2$. We find two different scaling behaviors 
depending on whether $x_0= O(1)$ (discrete quantum 
regime) or $x_0\sim n^{1/\mu}$ (classical scaling regime). Namely, 
to leading order one has (see Fig. \ref{Fig_crossover}),
\begin{equation}
q(x_0,n) \sim \left\{\begin{array}{rl}
\dfrac{1}{\sqrt{n}}\, U(x_0)   \, &\textrm{for}\,\,\quad  n\rightarrow +\infty\,\quad \textrm{and}\,\,\quad  x_0= O(1) \;,
    \\
\vspace{1.5mm}\\
V_\mu\left(\dfrac{x_0}{n^{1/\mu}}\right)  &\textrm{for}\,\,\quad n\rightarrow +\infty\,\quad \textrm{and}\,\,\quad x_0= 
O(n^{1/\mu}) \;.
\end{array}\right.
\label{result.1}
\end{equation}
The function $U(x_0)$ is given by its Laplace transform
\begin{equation}
\int_0^{\infty} U(x_0)\, e^{-\lambda\, x_0}\, dx_0= 
\frac{1}{\lambda\,\sqrt{\pi}}\, \exp\left[- 
\frac{\lambda}{\pi}\int_0^{\infty} \frac{ dk}{\lambda^2+k^2}\, 
\ln\left(1- {\hat f}(k)\right)\right]\, ,
\label{U.1}
\end{equation}
where ${\hat f}(k)$ is the Fourier transform of the jump PDF $f(\eta)$.
Thus, $U(x_0)$ depends on the full ${\hat f}(k)$ and not just
on its small-$k$ expansion (\ref{smallk.1}). For notational convenience, we will not make 
this dependence explicit and simply write $U(x_0)$ (instead of, e.g., $U_{\lbrack\hat{f}\rbrack}(x_0)$). 
The scaling function $V_{\mu}(z)$, by contrast, depends only on the small 
$k$ behavior of ${\hat f}(k)$ in Eq. (\ref{smallk.1}) and hence can be labelled just
by the index $\mu$. (Note that $U(x_0)$ cannot be labelled just by $\mu$
as it depends on the full $\hat f(k)$ and not just its small-$k$ behavior). 
We show that $V_\mu(z)$ is given by the following double integral 
transform
\begin{equation}
\int_0^{\infty} dy\, e^{-y}\, y^{1/\mu}\, \int_0^{\infty} dz\, V_\mu(z)\, 
e^{-w\, y^{1/\mu}\,z} = \frac{1}{w}\,J_\mu(w)\;,\quad {\rm where}\,\, 
J_\mu(w)= \exp\left[-\frac{1}{\pi}\int_0^{\infty} \frac{du}{1+u^2}\, 
\ln\left(1+ (a_\mu\, w\, u)^{\mu}\right)\right]\, .
\label{Vmu.1}
\end{equation}

While it is not easy to invert these integral transforms for a generic
jump PDF $f(\eta)$ (except in some special cases, see later), it is possible to
derive the large and small argument asymptotic behaviors of $U(x_0)$ and $V_\mu(z)$ explicitly.
For $U(x_0)$, we find
\begin{equation}
U(x_0) \sim \left\{\begin{array}{rl}
& \alpha_0 + \alpha_1\, x_0 + O(x_0^2)  \hskip 2.7cm \textrm{as}\,\, 
x_0\to 0
    \\
\vspace{1.5mm}\\
& A_\mu\, x_0^{\mu/2} + T_{\mu}(x_0)  
\quad\, \hspace*{2.6cm} \textrm{as} \,\, x_0\to \infty \;,
\end{array}\right.
\label{U_asymp.1}
\end{equation}
where $T_{\mu}(x_0) = o(x_0^{\mu/2})$ as $x_0 \to \infty$. The constants in Eq.\ (\ref{U_asymp.1}) are explicitly given by
\begin{eqnarray}
\alpha_0 & = & \frac{1}{\sqrt{\pi}} \label{alpha_0} \;, \\
\alpha_1 & =  & -\frac{1}{\pi^{3/2}}\, \int_0^{\infty} dk\, \ln\left(1- 
{\hat 
f}(k)\right) \label{alpha_1} \;, \\
A_\mu & = & \frac{a_\mu^{-\mu/2}}{\sqrt{\pi}\, \Gamma(1+\mu/2)} 
\label{Amu} \;. 
\end{eqnarray}
The large $x_0$ behavior of $T_{\mu}(x_0)$ is rather difficult to analyze explicitly for an arbitrary 
value of $\mu$, as it depends on the small-$k$ behavior of $\hat f(k)$ beyond the leading 
order (\ref{smallk.1}), involving higher order corrections $\hat f(k) = 1 - (a_\mu |k|)^\mu + c\,(a_\mu |k|)^\alpha + \ldots \;$, with $\alpha > \mu$.
However, in the case where $\hat f(k)$ is an analytic function at $k=0$, i.e. $\hat f(k) = 1 - (a_2 k)^2 + O(k^4)$ (corresponding thus to $\mu=2$ and $\alpha = 4$), one can show that $T_2(x_0) \to B_2$ as $x_0 \to \infty$, where $B_2$ is a computable constant. In this case, the second line of Eq. (\ref{U_asymp.1}) reduces to 
\begin{equation}
U(x_0)\sim A_2\, x_0 + B_2 + O(1/x_0) \quad {\rm as}\quad x_0\to \infty
\label{Uxmu2}
\end{equation}
with $A_2= 1/{a_2\sqrt{\pi}}$ and
\begin{equation}
B_2= - \frac{1}{a_2\,\pi^{3/2}}\,
\int_0^{\infty}     
\frac{dk}{k^2}\, \ln\left[\frac{1-{\hat f}(k)}{(a_2\, k)^{2}}\right]\, ,
\label{B2}
\end{equation}
which can be rewritten as
\begin{equation}
U(x_0)\sim \frac{1}{a_2\, \sqrt{\pi}}\left[x_0 + C_2\right]  
\quad {\rm as}\quad x_0\to \infty
\label{Uxmu2.1} 
\end{equation}
with
\begin{equation}
C_2= a_2\sqrt{\pi}\, B_2=-\frac{1}{\pi}\,\int_0^{\infty}
\frac{dk}{k^2}\, \ln\left[\frac{1-{\hat f}(k)}{(a_2\, 
k)^{2}}\right]\, .
\label{C2}
\end{equation}
Interestingly, this same constant $C_2$
has also appeared in a number of other contexts before, such as in the 
correction
term to the expected maximum of a random walk~\cite{CM2005,flajolet}, in the 
Smoluchowski trapping 
problem for Rayleigh flights in three dimensions~\cite{MCZ2006,ZMC2007,Ziff91} and as the so called ''Hopf constant'' in the physics of radiative transfer \cite{Milne,PS47,finch,Ivanov.1}.
For a discussion of this constant in another interesting context, see 
subsection \ref{sec:U_asympt}.

We now consider the scaling function $V_{\mu}(z)$ for $0<\mu\le 2$.
It turns out that for $\mu=2$, one can derive the scaling function
$V_2(z)$ exactly for all $z$. One finds
\begin{equation}
V_2(z)= {\rm erf}\left(\frac{z}{2\,a_2}\right)\, ,
\label{V2z.1}
\end{equation}
which is consistent with the Brownian result in Eq. (\ref{erf.1}) once
$2\, D\, t$ is replaced with $\sigma^2\, n= a_2^2\, n/2$, as expected for $\mu=2$. From Eq. (\ref{V2z.1}) one gets the asymptotic behaviors
\begin{equation}
V_2(z) \sim \left\{\begin{array}{rl}
 \dfrac{1}{a_2\sqrt{\pi}}\, z  \hskip 1.7cm &\textrm{as}\,\, z\to 0
    \\
\vspace{1.5mm}\\
 1- \dfrac{2\,a_2}{\sqrt{\pi} z}\, e^{-z^2/{4\,a_2^2}}
\quad\, &\textrm{as} \,\, z\to \infty \;.
\end{array}\right.
\label{V2_asymp.1}
\end{equation}

For $0<\mu<2$, 
the analysis of the scaling function $V_{\mu}(z)$ is
a little bit more complicated. Here, we provide the dominant asymptotic behaviors only,
\begin{equation}
V_\mu(z) \sim \left\{\begin{array}{ll}
 A_\mu\, z^{\mu/2}  \hskip 1.7cm &\textrm{as}\,\, z\to 0
    \\
\vspace{1.5mm}\\
 1- {\tilde A}_\mu\, z^{-\mu} 
\hskip 1.7cm &\textrm{as} \,\, z\to \infty \;,
\end{array}\right.
\label{Vmu_asymp.1}
\end{equation}
where $A_\mu$ is the same as in Eq. (\ref{Amu}) and
\begin{equation}
{\tilde A}_\mu = \frac{a_\mu^{\mu}}{\pi}\, \Gamma(\mu)\, 
\sin\left(\frac{\mu\pi}{2}\right) 
\hskip 0.5cm \textrm{with}\,\, 0<\mu< 2 \, . 
\label{tildeAmu}
\end{equation}
Using the aforementioned connection between $q(x_0,n)$ and the cumulative 
distribution of the global maximum $x_{\max}$, it is possible to relate $V_\mu(z)$ with the PDF of 
the maximum of a stable process, $v_\mu(z)$, which has been well studied in the mathematical literature (see e.g. \cite{Kuz2013} and references therein). Indeed, in the scaling regime $n$ and $x\to +\infty$ with fixed $x/n^{1/\mu}$ one has ${\rm Prob.}(x_{\max}=x) \sim n^{-1/\mu} v_\mu(x/n^{1/\mu})$, which yields $V_\mu(z) = \int_0^z v_\mu(z')\, dz'$. Note in particular that for $\mu = 1$, the PDF $v_1(z)$ can be computed exactly \cite{Dar1956}, providing thus an explicit integral representation of the scaling function $V_1(z)$ in this special case
\begin{eqnarray}\label{exact_v1}
V_1(z) = \frac{1}{\pi}\int_0^z dz' \, \frac{1}{z'^{1/2}(1+z'^2)^{3/4}} \, \exp{\left(-\frac{1}{\pi}\int_0^{z'} dw\,\frac{\ln w}{1+w^2}\right)} \;.
\end{eqnarray}
It can be checked that the asymptotics given above in Eq. (\ref{Vmu_asymp.1}) are fully compatible with the known series expansion of $v_\mu(z)$~\cite{Kuz2013}.

Finally, it can be seen that the leading order behaviors of $q(x_0,n)$ in the two regimes 
considered in Eq. (\ref{result.1}) match smoothly near their common limit of 
validity. Indeed, taking the large $x_0$ limit in the inner regime ($x_0= O(1)$) where
$q(x_0,n)\approx U(x_0)/\sqrt{n}$ in Eq. (\ref{result.1}) and
using the large $x_0$ behavior of $U(x_0)$ in Eq. 
(\ref{U_asymp.1}),
one gets $q(x_0,n)\simeq A_\mu\, x_0^{\mu}/\sqrt{n}$. Similarly, taking the small $x_0/n^{1/\mu}$ limit in the outer regime ($x_0= O(n^{1/\mu}$)) where $q(x_0,n)\approx V_\mu(x_0/n^{1/\mu})$ in Eq.~(\ref{result.1}) and using the small $z$ behavior of $V_\mu(z)$ in Eq. (\ref{Vmu_asymp.1}), one obtains exactly
the same result, $q(x_0,n)\simeq A_\mu\, x_0^{\mu}/\sqrt{n}$, 
ensuring a smooth matching between the two scales.

\begin{figure}
\includegraphics[width = 0.6\linewidth]{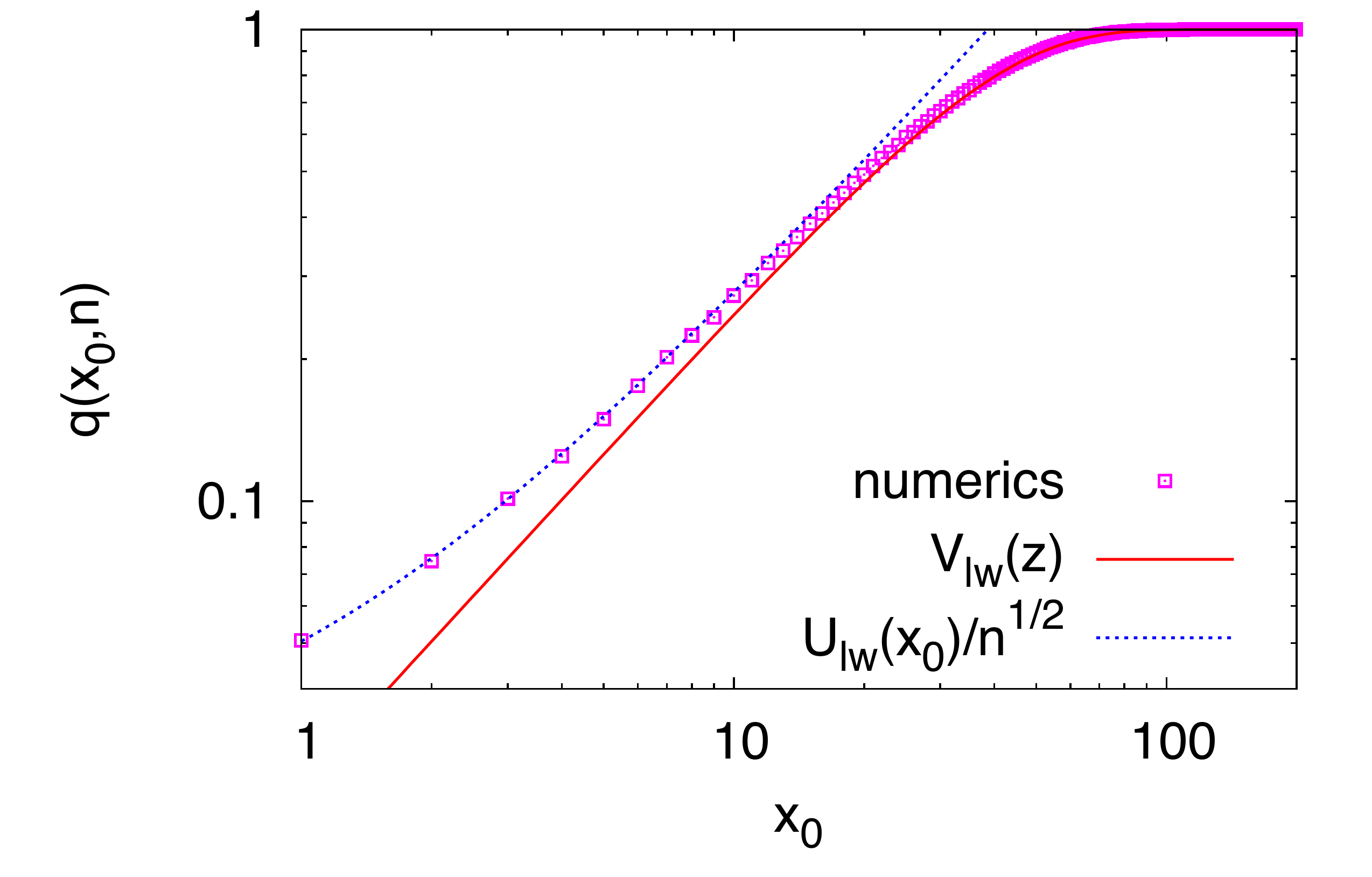}
\caption{{Log-log plot of $q(x_0,n)$ for the the lattice random walk (with lattice constant $1$), i.e. 
$f(\eta)= \frac{1}{2}\delta(\eta-1) + \frac{1}{2}\delta(\eta+1)$ and $n = 1000$ steps. The squares represent numerical results. The solid red line corresponds to $V_{\rm lw}(z = x_0/\sqrt{n})$ given in Eq. (\ref{v2z_lrw}) while the dotted blue line corresponds to $U_{\rm lw}(x_0)/\sqrt{n}$ with $U_{\rm lw}(x_0)$ given in Eq. (\ref{ux0_lrw_txt}). As for continuous jump distributions (see Fig. \ref{Fig_crossover}), this plot for the lattice random walk clearly illustrates the two different regimes for $x_0 = O(1)$ and $x_0 = O(\sqrt{n})$ as well as the crossover from the former to the latter as $x_0$ increases past $x_0\sim 30$.}}\label{Fig_crossover_lw}
\end{figure}

{{ Note that the main results mentioned above in Eqs. (\ref{result.1}) to (\ref{Vmu.1})
hold for symmetric and {\it continuous} jump
PDF $f(\eta)$ only. A natural example that does not belong to this class of continuous jump
densities is the lattice random walk (with lattice constant $1$) with $\pm 1$ jumps, i.e,
$f(\eta)= \frac{1}{2}\delta(\eta-1) + \frac{1}{2}\delta(\eta+1)$. While this
jump PDF is
symmetric, it is not continuous and hence our general formalism does not apply.
However, we show in Appendix B that in this special case, the survival probability $q(x_0,n)$ (where $x_0$ is a non-negative integer) can be worked out explicitly. One finds in particular that the asymptotic behavior
of $q(x_0,n)$ for large $n$ also has a `two scale' behavior (with $x_0\sim 
O(1)$ and 
$x_0\sim O(\sqrt{n}))$ as a function of $x_0$, similar to the continuous jump PDF case
in Eq. (\ref{result.1}), albeit with different scaling functions. More precisely, we show
in Appendix B  
that the counterpart of Eq. (\ref{result.1}) in the $\pm 1$ lattice random walk case is given by {(see Fig. \ref{Fig_crossover_lw})}
\begin{equation}  
q(x_0,n) \sim \left\{\begin{array}{rl}
\dfrac{1}{\sqrt{n}}\, U_{\rm lw}(x_0)   \, &\textrm{for}\,\,\quad
n\rightarrow +\infty\,\quad \textrm{and}\,\,\quad  x_0= O(1) \;,
    \\
\vspace{1.5mm}\\
V_{\rm lw}\left(\dfrac{x_0}{n^{1/2}}\right)
&\textrm{for}\,\,\quad n\rightarrow +\infty\,\quad \textrm{and}\,\,\quad
x_0=
O(n^{1/2}) \;,
\end{array}\right.
\label{result.lrw}
\end{equation}
with
\begin{eqnarray}
U_{\rm lw}(x_0) & =& \sqrt{\frac{2}{\pi}}\, (x_0+1) \quad {\rm for}\,\,
{\rm
all}\quad x_0\ge 0 \;,\label{ux0_lrw_txt} \\
V_{\rm lw}(z) & =& {\rm erf}\left(\frac{z}{\sqrt{2}}\right)\, ,
\label{v2z_lrw}
\end{eqnarray}
where the subscript '${\rm lw}$' stands for lattice walk. One can again check that
the behaviors of $q(x_0,n)$ match
smoothly at the boundary between the two regimes in Eq. (\ref{result.lrw}) (see Appendix-B for details).}

\section{Survival probability for discrete-time jump process: General 
Setting}

Consider the discrete-time jump process $x_n$ defined in Eq. 
(\ref{markov.1}) for a symmetric and continuous jump PDF 
$f(\eta)$. The survival probability $q(x_0,n)$, starting at $x_0\ge 0$, can be obtained from the `constrained' propagator $p_+(y,n|x_0)$ which is the PDF for the random walker, starting at $x_0$, to be at $y$ after $n$ steps, staying above 0 in between. It is easy to see that $p_+(y,n|x_0)$ satisfies the following backward master equation
\begin{eqnarray}\label{eq:p+}
p_+(y,n|x_0) = \int_0^\infty \, p_+(y,n-1|x_1) \,f(x_1-x_0)\,dx_1  \;,
\end{eqnarray}
with the initial condition $p_+(y,0|x_0) = \delta(y-x_0)$. This Eq. (\ref{eq:p+}) simply follows from the fact that in the first step, 
the walker jumps from $x_0\ge 0$ to some $x_1\ge 0$, staying positive, and 
then for the subsequent $(n-1)$ steps, the process renews with the new
initial position at $x_1$. Finally, one has to integrate over all
allowed positions $x_1\ge 0$ where the walker can jump to in the first 
step. The initial condition follows immediately from the definition of $p_+(y,n|x_0)$. From the ``constrained'' propagator, the survival probability can be simply obtained as
\begin{eqnarray}\label{rel_q_p}
q(x_0,n) = \int_{0}^\infty p_+(y,n|x_0) \,dy \;. 
\end{eqnarray} 
By integrating Eq. (\ref{eq:p+}) over the final position $y$, one immediately obtains that the survival probability $q(x_0,n)$ also evolves via a backward master equation
\begin{equation}
q(x_0,n)= \int_0^{\infty} q(x_1,n-1) \, f(x_1-x_0)\, dx_1
\label{master.1}
\end{equation}
with the initial condition $q(x_0,n=0)=1$ for all $x_0\ge 0$. Note that this master equation (\ref{master.1}) can be directly obtained following the reasoning described below Eq. (\ref{eq:p+}) and without using the constrained propagator $p_+(y,n|x_0)$.  

The evolution equation (\ref{eq:p+}) is deceptively simple but 
actually very hard to solve due to the integration extending over the
semi-infinite interval $[0,\infty)$ only. It belongs to the
general class of Wiener-Hopf integral equations which are notoriously not easy to solve for
an arbitrary kernel $f(\eta)$. However, for the case where the kernel
$f(\eta)$ has the interpretation of a probability density (i.e., non-negative for all arguments and normalizable to unity), there exists a solution to this equation which is given semi-explicitly by the so-called Hopf-Ivanov formula \cite{Ivanov1994} (for a `user friendly' derivation, see  
the Appendix A of Ref. \cite{MMS2014}) 
\begin{eqnarray}\label{ivanov}
\int_0^\infty \,dx_0 \, \int_0^\infty \,dy \, \sum_{n=0}^\infty p_+(y,n|x_0) \, s^n e^{-\lambda \,x_0 - \lambda' \, y} = \frac{\phi(s,\lambda)\, \phi(s,\lambda')}{\lambda+\lambda'} \;,
\end{eqnarray}
with
\begin{eqnarray}\label{def_phi}
\phi(s,\lambda)= \exp\left[- \frac{\lambda}{\pi}\int_0^{\infty} 
\frac{\ln \left(1-s\, {\hat f}(k)\right)}{\lambda^2+k^2}\, dk\right]\, ,
\end{eqnarray}
where ${\hat f}(k)$ is the Fourier transform of the jump PDF in Eq. 
(\ref{ft.1}). Setting $\lambda'=0$ in Eq. (\ref{ivanov}) and using Eq. (\ref{rel_q_p}) one obtains a formula for the survival probability 
\begin{eqnarray}\label{before_ps}
\sum_{n=0}^{\infty} s^n\, \int_0^{\infty} q(x_0,n)\, e^{-\lambda\, x_0}\, 
dx_0= \frac{\phi(s,\lambda) \phi(s,0)}{\lambda} \;.
\end{eqnarray}
The value $\phi(s,0)$ can be easily obtained from the expression in Eq. (\ref{def_phi}) by performing a change of variable $k = \lambda\,q$. Taking then the limit $\lambda \to 0$, using $\hat f(0) = 1$, one obtains that $\phi(s,0) = 1/\sqrt{1-s}$. Hence finally, one arrives at the  
%
%
%
so-called Pollaczek-Spitzer formula for the survival probability~\cite{Pollaczek,Spitzer}
\begin{equation}
\sum_{n=0}^{\infty} s^n\, \int_0^{\infty} q(x_0,n)\, e^{-\lambda\, x_0}\, 
dx_0= \frac{1}{\lambda\, \sqrt{1-s}}\,\phi(s,\lambda) \;. 
\label{PS1}
\end{equation}
While the solution in Eq. (\ref{PS1}) is exact, it is only semi-explicit
in the sense that one needs to invert the Laplace transform as well
as the generating function to obtain $q(x_0,n)$ fully explicitly.
This is possible in some special cases only, e.g. for the 
exponential
jump PDF $f(\eta)= e^{-|\eta|}/2$~\cite{CM2005,MCZ2006} (see Section \ref{examples}
for other special cases).
To derive the asymptotic properties of $q(x_0,n)$ from Eq. (\ref{PS1})
is a nontrivial technical challenge which has been discussed 
in several articles
~\cite{CM2005,MCZ2006,ZMC2007,M2010,SM2012,MMS2013,WMS2012,MSW2012,GMS2016}.

A remarkable simplification occurs if the starting point is exactly at the
origin, i.e., $x_0=0$. By taking the limit $\lambda\to \infty$ in Eq. 
(\ref{PS1}), it is easy to see that
\begin{equation}
\sum_{n=0}^{\infty} q(0,n)\, s^n = \frac{1}{\sqrt{1-s}}\, .
\label{SA.1}
\end{equation}
Thus, amazingly, the dependence on the jump PDF $f(\eta)$ 
disappears totally if $x_0=0$ ! This beautiful result goes by the name of the Sparre Andersen
theorem~\cite{SA}. Equating powers of $s$ on both sides of Eq.\ (\ref{SA.1}), one gets
\begin{equation}
q(0,n)= {2n\choose n}\, 2^{-2n}\, \,, 
\label{SA.4}
\end{equation}
a result that is completely {\em universal}, i.e., independent
of $f(\eta)$, for all $n$. It follows in particular that, for large $n$,
\begin{equation}
q(0,n) \sim \frac{1}{\sqrt{\pi\, n}}\ \ \ \ \ {\rm for}\ n\rightarrow +\infty,
\label{SA.5}
\end{equation}
which, obviously, is also universal. Let us emphasize again that this 
asymptotic result holds for any continuous and symmetric $f(\eta)$, 
including L\'evy flights with $\mu<2$.

In the next two sections we derive the large $n$ 
behavior of $q(x_0,n)$
in the two different regimes $x_0 = O(1)$ and $x_0 = O(n^{1/\mu})$ respectively, using as a starting point the Pollaczek-Spitzer formula in Eq. (\ref{PS1}).

\section{The discrete `quantum' regime $\bm{x_0= O(1)}$}

Inverting Eq. (\ref{PS1}) formally by using
Cauchy's integral formula, one gets
\begin{equation}
\int_0^{\infty} q(x_0,n)\, e^{-\lambda\,x_0}\, dx_0= \frac{1}{\lambda}\,
\int_C \frac{ds}{2\pi i}\, \frac{1}{s^{n+1}}\, 
\frac{\phi(s,\lambda)}{\sqrt{1-s}}\, ,
\label{cauchy.1}
\end{equation}
where the contour $C$ goes counter-clockwise around 
$s=0$ in the complex $s$-plane. From the expression of $\phi(s,\lambda)$ in Eq. (\ref{PS1}), it can be checked that if $\lambda\ne 0$, then $\phi(s,\lambda)$ is an analytic function of $s$ in the domain $\vert s\vert <1+\varepsilon$ for some $\varepsilon>0$ and $C$ can be deformed into a keyhole contour $C^\prime$ around the branch cut at $s=1$. For large $n$, the integral is dominated by the contribution of $C^\prime$ near $s=1$ and Eq.\ (\ref{cauchy.1}) reduces to
\begin{equation}
\int_0^{\infty} q(x_0,n)\, e^{-\lambda\,x_0}\, dx_0\sim 
\frac{\phi(1,\lambda)}{\lambda}\, \int_{C^\prime} \frac{ds}{2\pi i} 
\frac{1}{s^{n+1}}\, \frac{1}{\sqrt{1-s}}\ \ \ \ \ {\rm for}\ n\rightarrow +\infty\, . 
\label{cauchy.2}
\end{equation}
The $s$-integral in Eq. (\ref{cauchy.2}) is easily done by expanding $1/\sqrt{1-s}$ in a power series of $s$, applying the residue theorem for fixed $n$, and taking the limit $n\rightarrow +\infty$ in the result. One finds
\begin{equation}
\int_0^{\infty} q(x_0,n)\, e^{-\lambda\,x_0}\, dx_0\sim
\frac{1}{\sqrt{n}}\,\frac{\phi(1,\lambda)}{\lambda\, 
\sqrt{\pi}}\ \ \ \ \ {\rm for}\ n\rightarrow +\infty ,
\label{qx0_largen.1}
\end{equation}
and, by inverse Laplace transform,
\begin{equation}
q(x_0,n) \sim \frac{1}{\sqrt{n}}\, U(x_0)\ \ \ \ \ {\rm for}\ n\rightarrow +\infty\ {\rm at\ fixed}\ x_0 ,
\label{qx0_largen.2}
\end{equation}
where $U(x_0)$ is defined by its Laplace transform
\begin{equation}
\int_0^{\infty} U(x_0)\, e^{-\lambda\, x_0}\, dx_0= 
\frac{\phi(1,\lambda)}{\lambda\, {\sqrt \pi}}= 
\frac{1}{\lambda\,\sqrt{\pi}}\, \exp\left[-
\frac{\lambda}{\pi}\int_0^{\infty} \frac{ dk}{\lambda^2+k^2}\,
\ln\left(1- {\hat f}(k)\right)\right]\, .
\label{U.2}
\end{equation}
Here, we have used the expression of $\phi(s=1,\lambda)$ from Eq. (\ref{PS1}), which clearly shows that the function $U(x_0)$ depends on the full functional form of the jump PDF $f(\eta)$ through its Fourier transform ${\hat f}(k)$. It turns out that this function $U(x_0)$ satisfies a homogeneous integral equation,
\begin{equation}
U(x_0)= \int_0^{\infty} U(x_1)\, f(x_1-x_0)\, dx_1,
\label{homo.1}
\end{equation}
which can be shown by substituting the form $q(x_0,n)= 
U(x_0)/\sqrt{n}$ directly into the backward equation (\ref{master.1}).
For a given $f(\eta)$, the homogeneous equation (\ref{homo.1}) has a unique solution up to an 
overall multiplicative constant which can be fixed from the 
Sparre Andersen limit, $U(x_0=0)= 1/\sqrt{\pi}$.
Note however, that this solution is not that simple and is again given in terms of 
its 
Laplace transform only, as in Eq.~(\ref{U.2}).

\subsection{Asymptotics of $\bm{U(x_0)}$}\label{sec:U_asympt}

Except in some particular cases (see e.g. the examples in Sec.\ \ref{examples} below), it is generally not possible to determine the full expression of $U(x_0)$ explicitly. Nevertheless, it is always possible to extract the small and large $x_0$ asymptotic behaviors of $U(x_0)$ from Eq.\ (\ref{U.2}), as we will now show.

\subsubsection{The limit $x_0\to 0$}

The small $x_0$ behavior of 
$U(x_0)$ is given by the large $\lambda$ 
limit of the right-hand side (rhs) of Eq. (\ref{U.2}). Expanding for large
$\lambda$, one gets
\begin{eqnarray}
\int_0^{\infty} U(x_0)\, e^{-\lambda\, x_0}\, dx_0 &=& \frac{1}{\lambda 
\sqrt{\pi}}\, \exp\left[-\frac{1}{\pi \lambda}\, \int_0^{\infty} 
\ln\left(1-{\hat f}(k)\right)\, dk + 
O\left(\frac{1}{\lambda^2}\right)\right] \nonumber \\
&=& \frac{1}{\lambda
\sqrt{\pi}} - \frac{1}{\pi^{3/2} \lambda^2}\, \int_0^{\infty}
\ln\left(1-{\hat f}(k)\right)\, dk +
O\left(\frac{1}{\lambda^3}\right)\, ,
\label{ux_small.1}
\end{eqnarray}
and by inverse Laplace transform, one finds the following small $x_0$ behavior
\begin{equation}
U(x_0)= \alpha_0 + \alpha_1\, x_0 + O(x_0^2)
\label{ux_small.2}
\end{equation}
where the constants $\alpha_0$ and $\alpha_1$ are given in 
Eqs. (\ref{alpha_0}) and (\ref{alpha_1}) respectively. Note that Eqs.\ (\ref{qx0_largen.2}) and\ (\ref{ux_small.2}) yield $q(x_0=0,n)\sim\alpha_0/\sqrt{n}$ ($n\rightarrow +\infty$) with
$\alpha_0=1/\sqrt{\pi}$ independent of 
$f(\eta)$, which coincides precisely with 
the universal Sparre Andersen limit in Eq. (\ref{SA.5}), as it should be. 

\subsubsection{The limit $x_0\to \infty$}

The large $x_0$ behavior of 
$U(x_0)$ is given by the small $\lambda$ 
limit of the rhs of Eq. (\ref{U.2}), which is a bit more subtle to analyze.
Before letting $\lambda$ go to zero, first we need to separate out the most 
singular term. To this end, we note that 
${\hat f}(k) \sim 1- (a_\mu\, |k|)^{\mu}$ as $k\to 0$, and we write
\begin{equation}
\ln\left[1-{\hat f}(k)\right]= \ln\left[ \frac{1-{\hat f}(k)}{(a_\mu\, 
k)^{\mu}}\, (a_\mu\, k)^\mu\right]= \mu\, \ln (a_\mu\, k)+ 
\ln\left[\frac{1-{\hat f}(k)}{(a_\mu\,    
k)^{\mu}}\right]\, ,
\label{ux_large.1}
\end{equation}
on the rhs of Eq. (\ref{U.2}). Using then the identity
\begin{equation}
\int_0^{\infty} \frac{dk}{\lambda^2+k^2}\, \ln (a_\mu\, k)= 
\frac{\pi}{2\lambda}\, \ln (a_\mu\, \lambda)\; ,
\label{identity.1}
\end{equation}
we get
\begin{equation}
\int_0^{\infty} U(x_0)\, e^{-\lambda\, x_0}\, dx_0 =
\frac{1}{\sqrt{\pi}\, a_\mu^{\mu/2}\, \lambda^{1+\mu/2}}\, \exp\left[-K(\lambda)\right] \; ; 
\ {\rm where}\; K(\lambda) = \frac{\lambda}{\pi} \int_0^{\infty} \frac{dk}{\lambda^2+k^2} 
\ln\left[\frac{1-{\hat f}(k)}{(a_\mu\,
k)^{\mu}}\right] \;. \label{ux_large.2}
\end{equation}
Note that, so far, we have not taken the $\lambda\to 0$ limit. We have just 
re-written the rhs of Eq. (\ref{U.2}) in a way that will make it possible to control this
$\lambda\to 0$ limit in Eq.\ (\ref{ux_large.2}). In particular, one can show that $K(\lambda)=O(\lambda^{\zeta(\mu)})$
as $\lambda \to 0$, with $0<\zeta(\mu)\le 1$. Thus, for small $\lambda$ one has
\begin{eqnarray}
\int_0^{\infty} U(x_0)\, e^{-\lambda\, x_0}\, dx_0 = \frac{1}{\sqrt{\pi}\, a_\mu^{\mu/2}\, \lambda^{1+\mu/2}}\left\lbrack 1+O(\lambda^{\zeta(\mu)})\right\rbrack\; ,
\label{ux_large.3}
\end{eqnarray}
%
%
and by inverse Laplace transform, one obtains the large $x_0$ behavior
\begin{equation}
U(x_0) = A_\mu\, x_0^{\mu/2} + T_{\mu}(x_0)\;,
\label{ux_large.4}
\end{equation}
as announced in the second line of Eq. (\ref{U_asymp.1}), with
$T_{\mu}(x_0)=o(x_0^{\mu/2})$ for large $x_0$, and where the constant $A_\mu$ is given 
in Eq. (\ref{Amu}).
The analysis of the function $T_{\mu}(x_0)$ is more complicated as it depends on the small 
$k$ behavior (\ref{smallk.1}) of $\hat f(k)$, but also on the exponent $\alpha$ associated 
with the first correction to this leading behavior $\hat f(k) \approx 1 - (a_\mu |k|)^\mu + 
c \,  (a_\mu |k|)^\alpha$. We omit these details here, though they are 
straightforward to compute. 

Instead we focus on the case where $\hat f(k)$ is analytic at $k=0$, so that its small-$k$ 
expansion reads 
\begin{eqnarray}\label{f_analytics}
\hat f(k) = 1 -  (a_2\, k)^2 + c \,  (a_2 k)^4 + \cdots \;.
\end{eqnarray}
We start from Eq. (\ref{ux_large.2}) which is valid for general $\hat f(k)$. Setting $\mu = 2$ and using Eq. (\ref{f_analytics}) it is to see that, to leading order as $\lambda \to 0$, 
\begin{eqnarray}\label{K_small_l1}
K(\lambda) \to \frac{\lambda}{\pi} \int_0^\infty \frac{dk}{k^2} \, \ln{\left[\frac{1-\hat f(k)}{(a_2 \,k)^2}\right]} \;.
\end{eqnarray}
Note that this integral is convergent both in the infrared and the ultraviolet regimes. Therefore, keeping only the two leading terms in Eq. (\ref{ux_large.2}) gives
\begin{eqnarray}\label{K_small_l2}
\int_0^\infty U(x_0) \,e^{-\lambda x_0} \, dx_0 \sim \frac{1}{\sqrt{\pi}\, a_2\, \lambda^2}\left(1 - \frac{\lambda}{\pi} \int_0^\infty \frac{dk}{k^2} \, \ln{\left[\frac{1-\hat f(k)}{(a_2 \,k)^2}\right]}\right) \;.
\end{eqnarray}
Inverting this Laplace transform term by term, we get
\begin{equation}
U(x_0)\sim A_2\, x_0 + B_2 + O(1/x_0) \quad {\rm as}\quad x_0\to \infty
\label{Uxmu2_2}
\end{equation}
with 
\begin{equation}
A_2= 1/{a_2\sqrt{\pi}} \;, \quad\quad B_2= - \frac{1}{a_2\,\pi^{3/2}}\,
\int_0^{\infty}     
\frac{dk}{k^2}\, \ln\left[\frac{1-{\hat f}(k)}{(a_2\, k)^{2}}\right]\;.
\label{B2_2}
\end{equation}
Let us rewrite Eq. (\ref{Uxmu2_2}), using the expressions of 
$A_2$ and $B_2$ from Eq. (\ref{B2_2}), as
\begin{equation}
U(x_0)\sim \frac{1}{a_2\, \sqrt{\pi}}\left[x_0 + C_2\right]  
\quad {\rm as}\quad x_0\to \infty
\label{Uxmu2.1_2} 
\end{equation}
with
\begin{equation}
C_2= a_2\sqrt{\pi}\, B_2=-\frac{1}{\pi}\,\int_0^{\infty}
\frac{dk}{k^2}\, \ln\left[\frac{1-{\hat f}(k)}{(a_2\, 
k)^{2}}\right] \;.
\label{C2_2}
\end{equation}
This constant $C_2$ has the following interpretation. For $\hat f(k)$ that is analytic at $k=0$, 
the asymptotic behaviors of $U(x_0)$ are given by
\begin{equation}
U(x_0) \sim \left\{\begin{array}{rl}
& \frac{1}{\sqrt{\pi}} + O(x_0)  \hskip 2.7cm \textrm{as}\,\, 
x_0\to 0
    \\
\vspace{1.5mm}\\
&\frac{1}{a_2\, \sqrt{\pi}}\left[x_0 + C_2\right]    
\quad\, \hspace*{2.6cm} \textrm{as} \,\, x_0\to \infty \;,
\end{array}\right.
\label{U_asymp.1_mu2}
\end{equation}
with $C_2$ given in Eq. (\ref{C2_2}). The function $U(x_0)$, when plotted versus $x_0$, is 
asymptotically linear for large $x_0$. If we extrapolate this asymptotic large $x_0$ linear 
behavior all the way to $x_0<0$, it has a negative intercept at $x_0 = -C_2$ (see Fig. 
\ref{fig_hopf}). The survival probability $q(x_0,n) \sim U(x_0)/\sqrt{n}$, if extrapolated 
to negative $x_0$, vanishes at $x_0 = -C_2$. While the actual value $q(0,n) \sim 
1/\sqrt{\pi\,n}$ is nonzero at the origin $x_0=0$, the far-away profile instead indicates 
an effective location of the absorbing origin at $x_0 = -C_2$. In the context of the 
physics of radiative transfer, this constant $C_2$ is known as the Milne extrapolation 
length \cite{Milne,PS47}. This constant has also appeared in the Smoluchowski 
trapping problem for Rayleigh flights in three dimensions \cite{MCZ2006} and also in the subleading asymptotic large $n$ behavior of the expected maximum of a random walk of $n$ steps \cite{flajolet,CM2005,MCZ2006}. 

\begin{figure}
\includegraphics[width = 0.5\linewidth]{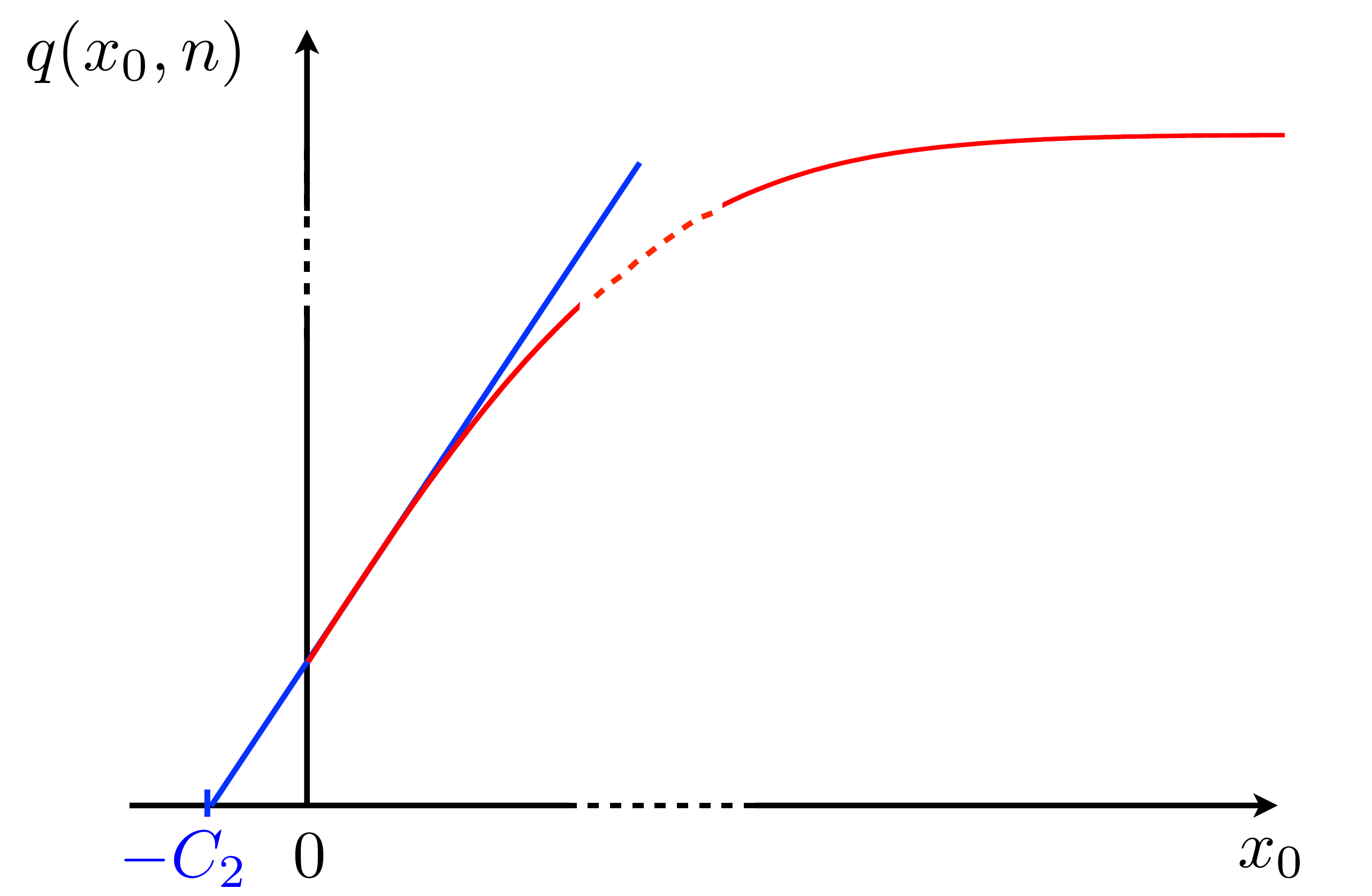}
\caption{Sketch of a plot of $q(x_0,n)$ (red line) as a function of $x_0$ (in arbitrary units along both axes). The blue line corresponds to $U(x_0)/\sqrt{n}$ as in Eq. (\ref{result.1}). The large $x_0$ behavior of $U(x_0)/\sqrt{n}$ [see Eq. (\ref{U_asymp.1_mu2})], when extrapolated to negative $x_0$, vanishes at $x_0=-C_2$.}\label{fig_hopf}
\end{figure}

In the context of radiative transfer of photons, the appropriate jump distribution turns out to
be \cite{Milne,PS47,Ivanov.1}
\begin{equation}
f(\eta)= \frac{1}{2}\, \int_0^1 \frac{du}{u}\, e^{-|\eta|/u} \;,
\label{ei.1}
\end{equation}
which is known as the Milne kernel. For this particular kernel, the constant $C_2$ came to be known as the Hopf constant $C_2^{\rm Hopf}$ \cite{finch}. Calculating $C_2^{\rm Hopf}$ has a long history, going back to Hopf who first computed it numerically, finding $C_2^{\rm Hopf}\approx 0.710$.
A decade later a team of physicists working in the Manhattan project
needed a more accurate value to find the critical mass of 
Uranium~\cite{Ivanov.1}, and $C_2^{\rm Hopf}$ was then computed to many 
decimal places. There indeed 
exists an analytical expression for this constant in terms of the so called Chandrasekhar 
$H$-function defined as \cite{Ivanov.2,stibbs}
\begin{eqnarray}\label{H_function}  
H(z) = \exp{\left[-\frac{z}{\pi} \int_0^\infty \ln{\left(1 - \frac{{\tan^{-1} 
\lambda) }}{\lambda} \right)} \frac{d\lambda}{1+z^2\lambda^2} \right]} \;.
\end{eqnarray}
The Hopf constant $C_2^{\rm Hopf}$ is then given by \cite{finch}
\begin{eqnarray}\label{Hopf_H}
C_2^{\rm Hopf} = \frac{\sqrt{3}}{2} \int_0^1 H(z)\,z^2 \, dz  = 0.7104460895 \ldots\;. 
\end{eqnarray}
{{ Using
a different method, Placzek and Seidel give an alternative and more compact analytical expression for the Hopf constant~\cite{PS47},
\begin{eqnarray}
C_2^{\rm Hopf}= \frac{6}{\pi^2} + \frac{1}{\pi}\, \int_0^{\pi/2} \left[\frac{3}{x^2}- 
\frac{1}{1-x\cot x}\right]\, dx= 0.7104460895 \ldots\;.
\label{PS_Hopf}
\end{eqnarray}
Yet another analytical expression for $C_2^{\rm Hopf}$ can be derived from our Eq. (\ref{C2_2}).}} Using the Milne jump
distribution in (\ref{ei.1}), we find that its Fourier transform is given by  
\begin{eqnarray}\label{Milne_fourier}
{\hat f}(k)=\frac{\tan^{-1} k}{k} \;,
\end{eqnarray}
which, for small-$k$,
behaves as ${\hat f}(k)= 1- k^2/3+\cdots$,
giving $a_2= 1/\sqrt{3}$ [see Eq.\ (\ref{smallk.1})]. Plugging this in Eq. (\ref{C2_2}), we get the following expression for the Hopf 
constant
\begin{equation}
C_2^{\rm Hopf}\, = -\frac{1}{\pi} \int_0^{\infty} 
\frac{dk}{k^2}\, \ln\left[\frac{3}{k^2}\left(1- 
\frac{\tan^{-1} k}{k}\right)\right]= 0.7104460895\ldots \;.
\label{Hopf_constant}
\end{equation}
We remark that in the notation of this paper, the Chandrasekhar 
$H$-function can be expressed in terms of the function 
$\phi(s,\lambda)$ defined in Eq. (\ref{def_phi}). Using the expression of $\hat f(k)$ in Eq. (\ref{Milne_fourier}), it is easy to check that 
\begin{eqnarray}\label{H_phi}
H(z) = \phi(1,1/z) \;.
\end{eqnarray}  
{{Here, we seemingly have three completely different analytical expressions for the Hopf constant $C_2^{\rm Hopf}$, respectively in Eqs.
(\ref{Hopf_H}), (\ref{PS_Hopf}) and (\ref{Hopf_constant}). Proving that these analytical expressions are actually equivalent, which is not immediately evident, will be interesting in itself and as it might give an insight of possible links between the seemingly different methods used to derive these expressions.}}

\subsection{Two exactly solvable cases}\label{examples}
There are few special cases of jump PDF $f(\eta)$ for which the
function $U(x_0)$ can be obtained explicitly by inverting
Eq. (\ref{U.2}). Here, we mention two examples in which such an inversion can be done.

\vskip 0.4cm

\noindent {\em Example I:} Consider the symmetric 
exponential 
distribution
\begin{equation} 
f(\eta)= \frac{1}{2a}\, e^{-|\eta|/a}\, ,
\label{exp.1}
\end{equation}
where $a$ is the typical jump length. 
Its Fourier transform is a Lorentzian 
\begin{equation}
{\hat f}(k)= \frac{1}{1+a^2\, k^2}\, .
\label{exp.2}
\end{equation}
Substituting ${\hat f}(k)$ on the rhs of Eq. (\ref{U.2}), one gets
\begin{equation}
\int_{0}^{\infty} U(x_0)\, e^{-\lambda\, x_0}\, dx_0= \frac{1}{\lambda 
\sqrt{\pi}}\, \exp\left[-\frac{\lambda}{\pi}\int_0^{\infty} 
\frac{dk}{\lambda^2+k^2}\, \left(\ln(a^2\, k^2)- 
\ln(a^2\, k^2+1)\right)\right]\, 
.
\label{exp.3}
\end{equation}
The log-integrals appearing on the rhs of Eq. (\ref{exp.3}) can be 
performed explicitly using the following useful identity
\begin{equation}
\int_0^{\infty} \frac{dk}{\lambda^2+k^2} \, \ln(\alpha+ \beta\, k^2)= 
\frac{\pi}{\lambda} \ln \left(\sqrt{\alpha}+ \sqrt{\beta}\, 
\lambda\right)\, ,
\label{iden.2}
\end{equation}
valid for any $\alpha\ge 0$ and $\beta\ge 0$. One obtains
\begin{equation}
\int_{0}^{\infty} U(x_0)\, e^{-\lambda\, x_0}\, dx_0= 
\frac{1}{\sqrt{\pi}}\left(\frac{1}{\lambda} + \frac{1}{a\, 
\lambda^2}\right) \,,
\label{exp.4}
\end{equation}
which can now be trivially inverted to give
\begin{equation}
U(x_0)= \frac{1}{\sqrt{\pi}}\left(1+ \frac{x_0}{a}\right)\, .
\label{ux_exp}
\end{equation}
In Fig. \ref{Fig_crossover} a), we show a plot of $U(x_0)/\sqrt{n}$ with $U(x_0)$ given in
Eq. (\ref{ux_exp}), and compare it to numerical simulations. 

\vskip 0.4cm

\noindent {\em Example II:} Consider now the symmetric gamma distribution of the form
\begin{equation}
f(\eta)= \frac{3\,|\eta|}{2\, a^2}\, e^{-\sqrt{3}\, |\eta|/a} \;.
\label{xexp.1}
\end{equation}
The choice of the constants are such that the variance is $2 a^2$.
The Fourier transform of $f(\eta)$ can again be easily obtained
\begin{equation}
{\hat f}(k)= \left(1- \frac{k^2\,a^2}{3}\right)\, \left(1+ 
\frac{k^2\,a^2}{3}\right)^{-2} \;,
\label{xexp.2}
\end{equation}
which behaves, for small-$k$, as ${\hat f}(k)= 1- a^2\, k^2$ as in Eq. 
(\ref{smallk.1}) with $a_2=a$. Substituting ${\hat f}(k)$ on the rhs of
Eq. (\ref{U.2}), and using again the identity in Eq. (\ref{iden.2}), one finds
\begin{equation}
\int_{0}^{\infty} U(x_0)\, e^{-\lambda\, x_0}\, dx_0= 
\frac{1}{\sqrt{\pi}\, a\, \lambda^2\, 
(1+a\lambda/3)}{\left(1+\frac{a\,\lambda}{\sqrt{3}}\right)}^2 \, .
\label{xexp.3}
\end{equation}
Using now the following break-up into rational fractions
\begin{equation}
\frac{(1+ b\lambda)^2}{\lambda^2 (1+c\lambda)}= 
\frac{(b-c)^2}{1+c\lambda}+ \frac{2b-c}{\lambda} + \frac{1}{\lambda^2}\;,
\label{ratfrac.1}
\end{equation}
and Laplace inverting each term, one finally gets the explicit expression
\begin{equation}
U(x_0)= \frac{1}{\sqrt{\pi}}\,\left[\frac{2\sqrt{3}-1}{3}+ \frac{x_0}{a}+ 
\frac{(\sqrt{3}-1)^2}{3}\, e^{-3x_0/a}\right]\, .
\label{ux_xexp}
\end{equation}
{In Fig. \ref{Fig_crossover} b), we show a plot of $U(x_0)/\sqrt{n}$ with $U(x_0)$ given in
Eq. (\ref{ux_xexp}), and compare it to numerical simulations.}

Evidently, it can be checked that $U(x_0)$ in Eqs. (\ref{ux_exp}) and (\ref{ux_xexp}) do satisfy the general 
asymptotic behaviors detailed in Eq. (\ref{U_asymp.1}) (with $\mu=2$ and 
$a_2=a$). {Additional numerical simulations, not shown here, confirm these asymptotic behaviors for other jump distributions, e.g. for a uniform and symmetric jump distribution, for which the full function $U(x_0)$ can not be computed explicitly.}

\section{The `classical' scaling regime 
$\bm{x_0= O(n^{1/\mu})}$} 

We now consider the survival probability in the scaling regime
defined by $x_0\sim n^{1/\mu}$ and large $n$. This scaling limit was already 
investigated in the context of the
statistics of the number of records for multiple random walks and
a derivation of Eq. (\ref{Vmu.1}) can be found in the appendices of Ref. \cite{WMS2012}.
There, the scaling function $V_\mu(z)$ was analyzed in the large $z$ limit only. What we now need to understand the matching with the inner scale $x_0= O(1)$ is the opposite limit $z\to 0$.
In order to make this paper as  
self-contained as possible, we include a detailed
derivation of Eq. (\ref{Vmu.1}) below and then provide the asymptotics
of $V_\mu(z)$ both for $z\to 0$ and $z\to \infty$. 

In the scaling limit, we 
need to take the
limits $n\to \infty$ and $x_0\to \infty$ keeping the 
ratio $z=x_0/n^{1/\mu}$ fixed in Eq.~(\ref{PS1}). In terms of the
two conjugate variables $s$ and $\lambda$, this scaling limit
translates into taking the limits $s\to 1$ and $\lambda\to 0$
keeping the ratio $\lambda/(1-s)^{1/\mu}$ fixed. To proceed
we set $p=1-s$ in Eq. (\ref{PS1}) and define the scaling variable
$w= \lambda/p^{1/\mu}$. Let us first analyse the rhs of Eq. (\ref{PS1})
in this scaling limit, in particular the function $\phi(s=1-p,\lambda)$.
Making the change of variable $k= \lambda u$ in the integral, we get 
\begin{equation}
\phi(s=1-p,\lambda) =  \exp\left[-\frac{1}{\pi}\int_0^{\infty} 
\frac{du}{1+u^2}\, \ln\left(1-(1-p)\, {\hat f}(\lambda u)\right)\right]\, 
.
\label{philimit.1}
\end{equation}
For $\lambda\to 0$ we can use the small-$k$ expansion
of ${\hat f}(k)$ in Eq. (\ref{smallk.1}). Then, we write 
$\lambda$ in terms of the scaled variable, $\lambda= w\, p^{1/\mu}$,
where $w$ is held fixed and $p\to 0$. This gives, to leading order in the 
scaling limit,
\begin{equation}
\phi(s=1-p,\lambda= w\, p^{1/\mu}) \approx  \exp\left[-\frac{1}{\pi}\int_0^{\infty}
\frac{du}{1+u^2}\, \ln\left(p \left(1 + 
(a_\mu\, w\, u)^\mu\right)\right) \right]\, . 
\label{philimit.2}
\end{equation}
Separating the $\ln p$ term and doing the integration, one gets
\begin{equation} 
\phi(s=1-p,\lambda= w\, p^{1/\mu}) \approx \frac{1}{\sqrt p}\, J_\mu(w);
\quad {\rm where}\,\,
J_\mu(w)= \exp\left[-\frac{1}{\pi}\int_0^{\infty} \frac{du}{1+u^2}\,
\ln\left(1+ (a_\mu\, w\, u)^{\mu})\right)\right]\, ,
\label{philimit.3}
\end{equation}
which, once substituted on the rhs of Eq. (\ref{PS1}), yields the
scaling limit
\begin{equation}
\sum_{n=0}^{\infty} e^{-p\,n}\, \int_0^{\infty} dx_0\, q(x_0,n)\, 
e^{- w\, p^{1/\mu}\, x_0}\,
\sim \frac{1}{p^{1+1/\mu}}\, 
\frac{1}{w}\, 
J_\mu(w)\ \ \ \ \ {\rm for}\ p\rightarrow 0\ {\rm and\ fixed}\ w .
\label{slimit.1}
\end{equation}
In order that both sides of Eq.\ (\ref{slimit.1}) have the same scaling
$\sim p^{-(1+1/\mu)}$ as $p\to 0$,
it is clear that $q(x_0,n)$ must scale as
\begin{equation}
q(x_0,n) \sim V_\mu\left(\frac{x_0}{n^{1/\mu}}\right)
\ \ \ \ \ {\rm for}\ \quad n\rightarrow +\infty\ \quad{\rm and}\\ \quad x_0\sim n^{1/\mu} \;.
\label{qscaling.1}
\end{equation}
To see this, we substitute the
anticipated form\ (\ref{qscaling.1}) on the left-hand side (lhs) of Eq. (\ref{slimit.1}), we write
$y= p\, n$ and, in the limit $p\to 0$, we replace the sum over $n$ by an integral over
$y$. Rescaling then $x_0= z n^{1/\mu}= z y^{1/\mu}\, 
p^{-1/\mu}$ we find that the lhs of~Eq. (\ref{slimit.1}) scales like $p^{-(1+1/\mu)}$ for $p\to 0$, as expected.
Cancelling the factor $p^{-(1+1/\mu)}$ from both sides, gives
the scaling function $V_\mu(z)$ as
\begin{equation}
\int_0^{\infty} dy\, e^{-y}\, y^{1/\mu}\, \int_0^{\infty} dz\, V_\mu(z)\,
e^{-w\, y^{1/\mu}\,z} = \frac{1}{w}\,J_\mu(w);\quad {\rm where}\,\,
J_\mu(w)= \exp\left[-\frac{1}{\pi}\int_0^{\infty} \frac{du}{1+u^2}\,
\ln\left(1+ (a_\mu\, w\, u)^{\mu}\right)\right]\, ,
\label{Vmu.2}
\end{equation}
which is the main result of this section. Thus, the scaling function $V_\mu(z)$ depends on the jump PDF $f(\eta)$ 
only through the L\'evy index $0<\mu\le 2$.
We will now see that for $\mu=2$, the expression of $V_2(z)$ can be obtained explicitly for all $z$, while for $0<\mu<2$, only the large and small $z$
asymptotics of $V_\mu(z)$ can be obtained.

\subsection{The case $\bm{\mu=2}$}

For $\mu=2$, the integral over $u$ in the expression\ (\ref{Vmu.2}) of $J_2(w)$ can be 
performed exactly by using
the identity in Eq.~(\ref{iden.2}). One finds
\begin{equation}
\int_0^{\infty} dy\, e^{-y}\, y^{1/2}\, \int_0^{\infty} dz\, V_2(z)\,
e^{-w\, y^{1/2}\,z} = \frac{1}{w\, (1+a_2\, w)}\, .
\label{v2z.1}
\end{equation}
It is still not easy to invert Eq. (\ref{v2z.1}). However, we know 
that for $\mu=2$ and in this scaling regime, we must recover the 
Brownian
result for $q(x_0,n)$ as given in Eq. (\ref{erf.1}) provided one replaces
$2\,D\,t$ with $\sigma^2\, n= a_2^2 n/2$. This suggests
that
\begin{equation}
V_2(z)= {\rm erf}\left(\frac{z}{2\,a_2}\right)\, .
\label{v2z.2}
\end{equation}
It can now be checked by substituting Eq. (\ref{v2z.2}) on the 
lhs of Eq. (\ref{v2z.1}) and performing the integrals, that  
the explicit expression of $V_2(z)$ in Eq.\ (\ref{v2z.2}) does indeed satisfy Eq. (\ref{v2z.1}), which justifies this expression a posteriori.

\subsection{The case $\bm{0<\mu\le 2}$}


For general $0<\mu \leq 2$, it is hard to obtain $V_\mu(z)$ explicitly for all $z$ from Eq. (\ref{Vmu.2}), except for $\mu = 1$ [see Eq. (\ref{exact_v1})]. In fact, for general $\mu$, even extracting the asymptotic behavior of $V_\mu(z)$ (for small and large $z$) from Eq. (\ref{Vmu.2}) is far from trivial. Fortunately, this can be done, as we demonstrate it below.

\subsubsection{The limit $z\to 0$}

The small $z$ limit of $V_{\mu}(z)$ corresponds to the large $w$ limit in Eq. (\ref{Vmu.2}).
To extract the large $w$ behavior of $J_\mu(w)$ in Eq. (\ref{Vmu.2}), we rewrite it as
\begin{eqnarray}
J_\mu(w) &=& \exp\left[-\frac{1}{\pi}\int_0^{\infty} \frac{du}{1+u^2}\,
\ln \left((a_\mu\,w\,u)^{\mu}\,(1+ (a_\mu\,w\,u)^{-\mu}\right)\right] 
\nonumber \\
&=& \exp\left[-\frac{\mu}{\pi} \int_0^{\infty} \frac{du}{1+u^2}\, 
\ln(a_\mu \,w\,u)- \frac{1}{\pi}\int_0^{\infty} \frac{du}{1+u^2}\,
\ln \left( 1+ (a_\mu\,w\,u)^{-\mu}\right)\right]\;.
\label{Jmu.1}
\end{eqnarray}
The first integral on the rhs in Eq. (\ref{Jmu.1}) can be done 
exactly and gives simply $ (\pi/2)\, \ln(a_\mu\,w)$. Hence, we get 
\begin{equation}
J_\mu(w)= \frac{1}{(a_\mu \,w)^{\mu/2}}\, \exp\left[-I_\mu(a_\mu\, 
w)\right]; \quad {\rm where}\quad I_\mu(w)=\frac{1}{\pi}\,\int_0^{\infty} 
\frac{du}{1+u^2}\, \ln\left(1+ (w\,u)^{-\mu}\right)\, .
\label{Imu.1}
\end{equation}
Note that, so far, we have not taken the $w\to \infty$ limit and Eq. 
(\ref{Imu.1}) is exact for all $w$. The purpose of the above manipulation
was just to extract the leading singularity $w^{-\mu/2}$ of 
$J_\mu(w)$. It now
remains to determine the large $w$ behavior of $I_\mu(w)$.

A careful analysis of the large $w$ limit of the integral $I_\mu(w)$ in Eq. (\ref{Imu.1}) gives
the following 
asymptotic behaviors (see Appendix \ref{Appendix_large_I} for details)
\begin{equation}
I_\mu(w)= \left\{\begin{array}{rl}
&\dfrac{1}{\sin(\pi/\mu)\, w} +O(w^{-2}) 
\quad \hskip 2.cm \textrm{when}\,\, \; 1<\mu\le 2 \;,
    \\
\vspace{1.5mm}\\
&\dfrac{1}{\pi\, w}\, \ln (w) + O(w^{-1}) \quad \hskip 2.2cm 
\textrm{when}\,\,\;\hspace*{0cm} \mu=1 \;\,, \\
\vspace{1.5mm}\\
&\dfrac{1}{2\, \cos(\pi\,\mu/2)\, w^{\mu}} + O\left(w^{-{\rm 
min}(1,2\,\mu)}\right)\quad \textrm{when}\,\, 0<\mu<1 \;.
\end{array}\right.
\label{Imu_asymp}
\end{equation}
Using these results in Eq. (\ref{Imu.1}),
we find that for all $0<\mu\le 2$,
\begin{equation}
J_\mu(w) \sim \frac{1}{(a_\mu \,w)^{\mu/2}}
\ \ \ \ \ {\rm for}\ w\rightarrow +\infty \;.
\label{Jmu.2}
\end{equation} 
Hence, from Eq. (\ref{Vmu.2}) we get
\begin{equation}
\int_0^{\infty} dy\, e^{-y}\, y^{1/\mu}\, \int_0^{\infty} dz\, V_\mu(z)\,
e^{-w\, y^{1/\mu}\,z} 
\sim \frac{1}{a_\mu^{\mu/2}\, w^{1+\mu/2}} 
\ \ \ \ \ {\rm for}\ w\rightarrow +\infty \, .
\label{Vmu.3}
\end{equation}
In order for both sides of Eq. (\ref{Vmu.3}) to have the same scaling $w^{-1-\mu/2}$ 
as $w\rightarrow +\infty$, it turns out that $V_\mu(z)\sim \gamma_\mu\, z^{\mu/2}$
as $z\to 0$, to be verified a posteriori and the unknown prefactor
$\gamma_\mu$ to be determined. Indeed, substituting this anticipated behavior
on the lhs of Eq. (\ref{Vmu.3}) and performing the
integrals we find
\begin{equation}
\gamma_\mu \int_0^{\infty} dy e^{-y}\, y^{1/\mu}\,\int_0^{\infty} 
dz\, z^{\mu/2}\, e^{-w\, y^{1/\mu}\, z}= \frac{\gamma_\mu\, \sqrt{\pi}\, 
\Gamma(1+\mu/2)}{w^{1+\mu/2}}\, ,  
\label{Vmu.4}
\end{equation}
justifying this small $z$ behavior of $V_{\mu}(z)$. Cancelling $w^{-(1+\mu/2)}$ from both sides,
we see that $\gamma_\mu= A_\mu$, where $A_\mu$ is given in Eq. 
(\ref{Amu}). Hence, finally, for all $0<\mu\le 2$, the
asymptotic behavior of 
$V_\mu(z)$ for small $z$ is given by
\begin{equation}
V_\mu(z)\sim  A_\mu\, z^{\mu/2}\ {\rm for}\ z\rightarrow 0,
\quad {\rm where}\quad A_\mu= 
\frac{a_\mu^{-\mu/2}}{\sqrt{\pi}\, \Gamma(1+\mu/2)}\, ,
\label{Vmu.5}
\end{equation}
as announced in Eq. (\ref{Vmu_asymp.1}).

\subsubsection{The limit $z\to \infty$}

The large $z$ limit of $V_\mu(z)$ corresponds to the small $w$ limit in Eq. (\ref{Vmu.2}). Clearly, from the expression of $J_\mu(w)$ in Eq.~(\ref{Vmu.2}) one has $J_\mu(0) = 1$ and the large $z$ behavior of $V_\mu(z)$ is actually determined by the leading singular correction in $J_\mu(w)$ as $w \to 0$. This correction depends on whether $0<\mu<1$, $1< \mu<2$, or $\mu=1$. One finds (see \cite{WMS2012} for details)

\begin{eqnarray}\label{Jmu_small_w}
J_{\mu}(w) \sim
\begin{cases}
&1 - b_\mu \, w^\mu \;, \; 0<\mu<1 \;, \\
& \\
& 1 + \dfrac{a_1}{\pi} \, w\, \ln w\;,\; \mu = 1 \;, \\
& \\
& 1 - \alpha_\mu \, w - b_\mu \, w^{\mu}\;, \; 1<\mu\leq 2 \;,
\end{cases}
\ \ \ \ \ {\rm for}\ w\rightarrow 0
\end{eqnarray}   
where the amplitudes $\alpha_\mu$ and $b_\mu$ are given by~\cite{WMS2012}
\begin{eqnarray}\label{bmu}
\alpha_{\mu} = J_\mu'(0) = -\frac{a_\mu}{\sin{(\pi/\mu)}} \;\;, \;\; b_\mu = \frac{a_\mu^\mu}{2\cos{(\mu\pi/2)}} \;.
\end{eqnarray}
To get the large $z$ behavior of $V_{\mu}(z)$ from Eqs. (\ref{Vmu.2}) and (\ref{Jmu_small_w}) it is convenient to introduce its Laplace transform 
\begin{eqnarray}\label{Laplace_V}
\tilde V_\mu(\rho) = \int_0^\infty V_{\mu}(z) e^{-\rho\,z} \,dz \;,
\end{eqnarray}
so that the equation determining $V_\mu(z)$ in Eq. (\ref{Vmu.2}) reads
\begin{eqnarray}\label{Vmu_Laplace}
\int_0^\infty dy\, e^{-y} y^{1/\mu} \, \tilde V_{\mu}(w\,y^{1/\mu}) = \frac{1}{w}J_\mu(w) \;.
\end{eqnarray}
From this equation, and the small $w$ behavior of $J_\mu(w)$ in the first line of Eq. (\ref{Jmu_small_w}) one can determine the small $\rho$ behavior of $\tilde V_{\mu}(\rho)$. Again, the three cases $0<\mu<1$, $1< \mu<2$ and $\mu=1$ have to be analyzed separately. 

\bigskip
{\bf $\bm{0 < \mu < 1}$.} In this case, by inserting a power law behavior of $\tilde V_{\mu}(w\,y^{1/\mu})$, valid for small $w$, on the lhs of Eq. (\ref{Vmu_Laplace}) and using the small $w$ expansion of $J_\mu(w)$ (\ref{Jmu_small_w}) on the rhs of Eq. (\ref{Vmu_Laplace}), one obtains by matching the powers of $w$ on both sides:  
\begin{eqnarray}\label{tildeV_1}
\tilde V_\mu(\rho) \sim \frac{1}{\rho} - b_\mu \,\rho^{\mu-1}
\ \ \ \ \ {\rm for}\ \rho\rightarrow 0\;,
\end{eqnarray}
where $b_\mu$ is given in Eq. (\ref{bmu}). Using then a Tauberian theorem, one finds 
\begin{eqnarray}\label{V_largez_1}
V_\mu(z) \sim 1 - \tilde A_\mu \,z^{-\mu}\ {\rm for}\ z\rightarrow +\infty
\quad, \;\quad{\rm where} \;\quad \tilde A_{\mu} = \frac{b_\mu}{\Gamma(1-\mu)} = \frac{a_\mu^\mu}{2 \, \Gamma(1-\mu)\,\cos{\mu\pi/2}} \;.
\end{eqnarray}
Finally, using Euler's reflection formula $\Gamma(1-\mu) \Gamma(\mu) = \pi/\sin{(\pi \mu)}$ (valid for $\mu \notin {\mathbb Z}$) one obtains the result announced in Eq. (\ref{Vmu_asymp.1}). 

\bigskip
{\bf $\bm{1 < \mu < 2}$.} The same analysis can be done in this case, where the small $w$ behavior of $J_\mu(w)$ is now given by the third line of Eq. (\ref{Jmu_small_w}). We get
\begin{eqnarray}\label{tildeV_2}
\tilde V_\mu(\rho) \sim \frac{1}{\rho} - \frac{\alpha_\mu}{\Gamma(1+1/\mu)} - b_\mu \,\rho^{\mu-1}
\ \ \ \ \ {\rm for}\ \rho\rightarrow 0\;,
\end{eqnarray} 
where $\alpha_\mu$ and $b_\mu$ are given in Eq. (\ref{bmu}). The leading term in this expansion, i.e., $1/\rho$, is the same as the one obtained for $0<\mu<1$ in Eq. (\ref{tildeV_1}).
The first (regular) correction is a constant term associated with the regular part of the scaling function that decays rapidly for large $z$ (note that there was a typo in Eq. (B24) of Ref. \cite{WMS2012}). Its contribution to the large $z$ behavior is negligible compared to the one of the third term in Eq.\ (\ref{tildeV_2}) which is the first singular correction giving rise to the algebraic decay 
\begin{eqnarray}\label{V_largez_3}
V_\mu(z) \sim 1 - \tilde A_\mu\,z^{-\mu}
\ \ \ \ \ {\rm for}\ z\rightarrow +\infty\;, 
\end{eqnarray}
as announced in Eq. (\ref{Vmu_asymp.1}).

\bigskip
{\bf $\bm{\mu=1}$.} This case needs to be analyzed separately because of the presence of logarithmic corrections in the small $w$ behavior of $J_1(w)$ (see the second line of Eq. (\ref{Jmu_small_w})). By using this asymptotic behavior in Eq. (\ref{Vmu_Laplace}) we find that $\tilde V_1(\rho)$ behaves, for small $\rho$, as
\begin{eqnarray}\label{tildeV_3}
\tilde V_1(\rho) \sim \frac{1}{\rho} + \frac{a_1}{\pi} \ln \rho
\ \ \ \ \ {\rm for}\ \rho\rightarrow 0\;,
\end{eqnarray}    
from which one straightforwardly obtains the large $z$ behavior
\begin{eqnarray}\label{V_largez_2}
V_1(z) \sim 1 - \tilde A_1\,z^{-1}\ {\rm for}\ z\rightarrow +\infty
\quad, \;\quad{\rm where} \;\quad \tilde A_{1} = \frac{a_1}{\pi} \;,
\end{eqnarray} 
as announced in Eq. (\ref{Vmu_asymp.1}). 
 
 \bigskip
To conclude, we point out that, while the small $w$ behavior of $J_\mu(w)$ differs in the three cases $0<\mu<1$, $1< \mu<2$ or $\mu=1$, the functional dependence of the amplitude $\tilde A_\mu$ in Eq. (\ref{Vmu_asymp.1}) on the parameter $\mu$ is the same on the whole interval $\mu \in (0,2)$ (this fact was actually overlooked in Ref. \cite{WMS2012}).

\section{Conclusion}

In this paper we have investigated the persistence, or survival probability, for long 
random walks and L\'evy flights as a function of the starting position. Assuming that the 
walker starts at some $x_0\ge 0$, we have identified two different regimes depending on how 
$x_0$ scales with the number of steps $n$ in the walk. These two regimes determine the late 
time behavior of the survival probability as given by Eq.\ (\ref{result.1}).

The classical, or standard, scaling regime is defined by $x_0=O(n^{1/\mu})$ for large $n$. 
It is well-known that with the resolution of the standard rescaled variables, the 
discrete-time random walk appears as a continuous time random process for which the 
persistence goes to zero with the starting point [see Eq.\ (\ref{qxt_asymp})]. On the other 
hand, in the inner, or quantum, regime defined by $x_0=O(1)$, the persistence goes to a 
finite (non-zero) value as $x_0$ goes to zero, as given by the Sparre Andersen results in 
Eqs.\ (\ref{SA.2}) and\ (\ref{SA.3}). Note that with the resolution of the standard 
rescaled variables, this latter regime lives inside a thin boundary layer of width $\sim 
n^{-1/\mu}\ll 1$ located near (and above) the origin, hence the name of `inner' regime. 
(The fact that the discrete-time nature of the walk can no longer be neglected in this 
boundary layer justifies the alternative name of `quantum' regime we have also used). It is 
the very existence of this boundary layer, in which the quantum regime takes the place of 
the classical one, that lifts the apparent paradox between the classical and Sparre 
Andersen results [Eqs.\ (\ref{qxt_asymp}) and\ (\ref{SA.3}), respectively]. For any fixed, 
arbitrarily large, $n$ there is always a starting position $x_0>0$ (or $x_0/n^{1/\mu}>0$ in 
the rescaled variables) below which the second line in Eq.\ (\ref{result.1}) must be 
replaced with the first line, leading to the Sparre Andersen limit\ (\ref{SA.3}) which 
always gives the (unique) correct result for $x_0=0$.

By a careful analysis of the asymptotics of the functions $U(x_0)$ and $V_{\mu}(z)$ 
appearing on the rhs of Eq.\ (\ref{result.1}), we have proved that the classical and 
quantum regimes match smoothly near there common limit of validity, in the sense that the 
large $x_0$ limit of $U(x_0)$ in the first line of Eq.\ (\ref{result.1}) (inner regime) 
coincides with the small $x_0/n^{1/\mu}$ limit of $V_{\mu}(x_0/n^{1/\mu})$ in the second 
line of Eq.\ (\ref{result.1}) (standard scaling regime). Our analysis also shows that there 
are only two scales in the large $n$ behavior of $q(x_0,n)$, namely $x_0= O(1)$ and $x_0= 
O(n^{1/\mu})$.

Indeed, this kind of quantum to classical crossover, as a function of temperature, also 
occurs in real quantum problems.  For instance, for $N$ non-interacting fermions in a 
one-dimensional harmonic trap of frequency $\omega$, the statistics of the kinetic energy, 
as a function of temperature $T$, exhibits two scales \cite{jacek}: one when it is of the 
order of the energy gap between single particle levels ($k_B\,T \sim \hbar \omega$), and 
the other one when it is of the order of the Fermi energy $E_F$ ($k_B\, T \sim N \, \hbar 
\omega \sim E_F$). When $k_B \,T = O(\hbar \omega)$, quantum fluctuations dominate the 
statistics of the kinetic energy while for $k_B \,T = O(N \hbar \omega)$ the 
classical/thermal fluctuations take over. The crossover between the quantum and the 
classical regimes in this quantum problem is very much reminiscent, at least qualitatively, 
to the crossover described in this paper.

{{ As mentioned at the end of the introduction, while the spirit of this paper is that 
of a 
review, there are nevertheless a few new results that we have not seen in the 
published literature. For the reader's sake, we provide here a list
of these new results: (i) The existence of two scales of $x_0$ (the discrete `quantum' 
regime where $x_0\sim O(1)$ and the `classical' scaling regime where $x_0\sim 
O(n^{1/\mu})$) in the asymptotic behavior of $q(x_0,n)$ as a function of $x_0$,
as well as the demonstration of the smooth matching between these two different scales, is new 
to the best of our knowledge. (ii) The
results for $q(x_0,n)$ in Eqs. (\ref{result.1}), with
the expression for $U(x_0)$ in Eq. (\ref{U.1}) appear to be new. Similarly, the asymptotic 
properties of $U(x_0)$ in Eq. (\ref{U_asymp.1}), along with the general expressions
for the constants in Eqs. (\ref{alpha_0}), (\ref{alpha_1}) and (\ref{Amu}) are also
new.  (iii) The expression in Eq. (\ref{C2}) for the constant $C_2$, valid for arbitrary
symmetric and continuous jump PDF $f(\eta)$ is new, and for the particular case
of the Milne jump PDF in Eq. (\ref{ei.1}) our general result in Eq. (\ref{C2}) provides
a new analytical expression for the Hopf constant $C_2^{\rm Hopf}$ in Eq. (\ref{Hopf_constant})
that seems different from the other known expressions for this famous 
constant in the literature.}}

\appendix

\section{The large $\bm{w}$ behavior of $\bm{I_\mu(w)}$}\label{Appendix_large_I}

In this appendix, we analyse the large $w$ behavior of the following
integral
\begin{equation}
I_\mu(w)= \frac{1}{\pi}\int_0^{\infty} \frac{du}{1+u^2}\, \ln\left(1+ 
(wu)^{-\mu}\right)\, ; \quad\quad 0<\mu\le 2 \, .
\label{Iw.1}
\end{equation}

\vskip 0.3cm

\noindent {\bf {The case $\bm{1<\mu\le 2}$:}} We first partition the integral in 
Eq. (\ref{Iw.1}) into two separate integrals as follows 
\begin{eqnarray}
I_\mu(w) &= & \frac{1}{\pi} \int_0^{\infty} du\, \ln\left(1+ 
(wu)^{-\mu}\right)\left[1- \frac{u^2}{1+u^2}\right] \nonumber \\
& =& \frac{1}{\pi w}\int_0^{\infty} dq\, \ln(1+q^{-\mu})- \frac{1}{\pi} 
\int_0^{\infty} \frac{du\, u^2}{1+u^2}\, \ln\left(1+ 
(wu)^{-\mu}\right)= S_1(w) + S_2(w)
\label{Iw.11}
\end{eqnarray}
where in the first integral we have made a change of variable $wu=q$.
Note that the first integral $S_1(w)$ is convergent for $1<\mu\le 2$, and 
yields
the leading behavior as $w\to \infty$
\begin{equation}
S_1(w)= \frac{1}{\sin(\pi/\mu)\, w}\, .
\label{s1w}
\end{equation}
The second integral $S_2(w)$ provides only 
subleading corrections for large $w$. The first subleading term can be 
estimated by expanding the logarithm in the integrand in $S_2(w)$ and 
keeping the first term 
\begin{equation}
S_2(w)\approx - \frac{w^{-\mu}}{\pi}\int_0^{\infty} du 
\frac{u^{2-\mu}}{1+u^2}= \frac{w^{-\mu}}{2\, \cos(\pi\mu/2)}\,.
\label{s2w}
\end{equation}
Evidently, the integral is convergent for $1<\mu\le 2$.
Adding the two integrals then yields the leading large $w$ behavior of 
$I_\mu(w)$ for $1<\mu\le 2$
\begin{equation}
I_\mu(w) = \frac{1}{\sin(\pi/\mu)\, w} + O(w^{-\mu})\, , \quad\quad 
1<\mu\le 2\, .
\label{Iw.12}
\end{equation}

\vskip 0.3cm 

\noindent {\bf {The case $\bm{\mu=1}$:}} In this case
\begin{equation}
I_1(w)= \frac{1}{\pi}\int_0^{\infty} \frac{du}{1+u^2}\, \ln\left(1+ 
\frac{1}{wu}\right)\, .
\label{I1w.1}
\end{equation}
To extract the leading large $w$ behavior, it is convenient to first take
the derivative if $I_1(w)$ with respect to $w$. The resulting
integral for $I_1'(w)=dI_1(w)/dw$ can be performed exactly. We get
\begin{equation}
I_1'(w)= -\frac{1}{\pi w} \int_0^{\infty} 
\frac{du}{(1+u^2)(1+wu)}=-\frac{\pi+2w\ln(w)}{2\pi w(1+w^2)}\, .
\label{I1w.2}
\end{equation}
We now expand the rhs of Eq. (\ref{I1w.2}) for large $w$. This gives 
\begin{equation}
I_1'(w) = \frac{-\ln w}{\pi w^2}\left[1+ \frac{\pi}{2w\ln w}+ 
O(w^{-2})\right]\,.
\label{I1w.3}
\end{equation}
Integrating back, using $I_1(\infty)=0$, immediately gives the large $w$ 
asymptotics 
\begin{equation}
I_1(w) = \frac{\ln w}{\pi w} + O(w^{-1})\, , \quad\quad \mu=1\, .
\label{I1w.4}
\end{equation}

\vskip 0.3cm

\noindent {\bf {The case $\bm{0<\mu <1}$:}} In this case, we again separate 
the integral $I_\mu(w)$ in Eq. (\ref{Iw.1}) into two parts, but in a 
different way from in Eq. (\ref{Iw.11}), as follows
\begin{eqnarray}
I_\mu(w) &=& \frac{1}{\pi}\int_0^{\infty} \frac{du}{1+u^2} 
\left[\ln\left(1+ (wu)^{-\mu}\right)- (wu)^{-\mu}+ (wu)^{-\mu}\right] 
\nonumber \\
&=& \frac{w^{-\mu}}{\pi}\int_0^{\infty} \frac{du\, u^{-\mu}}{1+u^2} 
+\frac{1}{\pi}\int_0^{\infty} \frac{du}{1+u^2} \left[\ln\left(1+ 
(wu)^{-\mu}\right)- (wu)^{-\mu}\right]\nonumber \\
&=& T_1(w) + T_2(w) \, .
\label{Iwl.1}
\end{eqnarray}
The first integral $T_1(w)$ is convergent for all $0<\mu <1$ and yields the
leading behavior for large $w$
\begin{equation}
T_1(w)= \frac{w^{-\mu}}{\pi}\int_0^{\infty} \frac{du\, u^{-\mu}}{1+u^2}= 
\frac{1}{2\cos(\mu\pi/2)\, w^{\mu}}\, , \quad\quad 0<\mu<1\, .
\label{Iwl.2}
\end{equation}

The second integral $T_2(w)$ yields subleading terms for large $w$. To
extract the leading behavior of $T_2(w)$, we first make a change of 
variable $wu=q$ in the expression of $T_2(w)$ in Eq. (\ref{Iwl.1}). This 
gives
\begin{equation}
T_2(w)= \frac{w}{\pi} \int_0^{\infty} \frac{dq}{q^2+w^2}\left[ 
\ln(1+q^{-\mu})- q^{-\mu}\right]\, .
\label{T2w.1}
\end{equation}
We note that the term $\ln(1+q^{-\mu})- q^{-\mu}\sim q^{-2\mu}$ for large 
$q$. Thus, for large $w$, if we approximate $q^2+w^2\approx w^2$ in the 
denominator of the integrand in Eq. (\ref{T2w.1}), the resulting integral
remains convergent, provided $1/2<\mu<1$. Thus in the range $1/2<\mu<1$, we 
get, for large $w$, the leading behavior of $T_2(w)$ as
\begin{equation}
T_2(w)\approx \frac{1}{\pi w} \int_0^{\infty} dq\, \left[\ln(1+q^{-\mu})- 
q^{-\mu}\right] = O(1/w)\, ; \quad\quad 1/2<\mu<1 \, .
\label{T2w.2}
\end{equation}
If however $0<\mu<1/2$, the integral is not convergent and this 
approximation does not work. In this case, one can re-start from the 
expression of $T_2(w)$ in Eq. (\ref{T2w.1}) and re-write the integral
as
\begin{eqnarray}
T_2(w) &= & \frac{w}{\pi}\int_0^{\infty} \frac{dq}{q^2+w^2} 
\left[\ln(1+q^{-\mu})- q^{-\mu}+ 
\frac{1}{2}q^{-2\mu}-\frac{1}{2}q^{-2\mu}\right]\nonumber \\
&=& -\frac{w}{2\pi}\int_0^{\infty} \frac{dq\, q^{-2\mu}}{q^2+w^2} 
+\frac{w}{\pi}\int_0^{\infty} \frac{dq}{q^2+w^2}\left[\ln(1+q^{-\mu})- 
q^{-\mu}+\frac{1}{2}q^{-2\mu}\right]\nonumber \\
&=& T_{21}(w) + T_{22}(w)\, .
\label{T2w.3}
\end{eqnarray}
The first integral $T_{21}(w)$ is convergent for all $0<\mu<1/2$ and gives
after a change of variable $q=wv$
\begin{equation}
T_{21}(w)= -\frac{w^{-2\mu}}{2\pi}\int_0^{\infty} \frac{dv\, 
v^{-2\mu}}{1+v^2}=- \frac{w^{-2\mu}}{4 \cos(\mu \pi)}\, ; \quad\quad 
0<\mu<1/2 \;.
\label{T2w.4}
\end{equation}
One can then analyse the second integral $T_{22}(w)$ in Eq. (\ref{T2w.3}) 
in a similar way as before, namely, we can approximate $q^2+w^2\approx w^2$ 
for large $w$ in the denominator of $T_{22}(w)$. The resulting integral
is convergent, provided $1/3<\mu<1/2$. This gives
\begin{equation}
T_{22}(w)\approx \frac{1}{\pi w}\int_0^{\infty} dq 
\left[\ln(1+q^{-\mu})-q^{-\mu}- \frac{1}{2}q^{-2\mu}\right] =
O(\frac{1}{w})\, ; 
\quad \quad 1/3<\mu<1/2 \, .
\label{T2w.5}
\end{equation}
For $\mu<1/3$, one can again repeat the same procedure (of subtracting the
singular terms in the expansion of $\ln(1+q^{-\mu})$ for large $q$) and
it is easy to see that $T_{22}(w) \sim w^{-3\mu}$ for $1/4<\mu<1/3$ etc.
Thus, adding $T_{21}(w)$ and $T_{22}(w)$, we find that the leading order
behavior of $T_2(w)$ is given by
\begin{eqnarray}
T_2(w) &\sim& w^{-1}\; , \quad\,\,\, {\rm for}\quad 1/2<\mu<1 \nonumber \\
& \sim & w^{-2\mu}\; , \quad\, {\rm for}\quad     0<\mu<1/2 \, .
\label{T2w.6}
\end{eqnarray} 
The asymptotic large $w$ behaviors in Eq. (\ref{T2w.6}) can be put together 
for 
all $0<\mu<1$ simply as
\begin{equation}
T_2(w)\sim w^{-{\rm min}(1, 2\mu)}\, , 
\quad\quad 0<\mu<1\, .
\label{T2w.7}
\end{equation}
Finally, adding $T_1(w)$ in Eq. (\ref{Iwl.2}) and $T_2(w)$ in Eq. 
(\ref{T2w.7}), we get the large $w$ asymptotic behavior of $I_\mu(w)$
for $0<\mu<1$
\begin{equation}
I_\mu(w)= \frac{1}{2\cos(\mu\pi/2)\, w^{\mu}} + O\left(w^{-{\rm min}(1, 
2\mu)}\right)\, , \quad\quad 0<\mu<1\, .
\label{Iw_final}
\end{equation}
 
Summarizing, we find the following large $w$ asymptotics of $I_\mu(w)$ in
Eq. (\ref{Iw.1}) for all $0<\mu\le 2$
\begin{equation}
I_\mu(w)= \left\{\begin{array}{rl}
&\dfrac{1}{\sin(\pi/\mu)\, w} +O(w^{-\mu})
\quad \hskip 1.8cm \textrm{when}\,\, \; 1<\mu\le 2
    \\
\vspace{1.5mm}\\
&\dfrac{1}{\pi\, w}\, \ln (w) + O(w^{-1}) \quad \hskip 2.0cm
\textrm{when}\,\, \mu=1 \;, \\
\vspace{1.5mm}\\
&\dfrac{1}{2\, \cos(\pi\,\mu/2)\, w^{\mu}} + O\left(w^{-{\rm
min}(1,2\,\mu)}\right)\quad \textrm{when}\,\, 0<\mu<1 \;,
\end{array}\right.
\label{Imu_asymp_app}
\end{equation}
as given in Eq. (\ref{Imu_asymp}) in the text.

\section{ Survival probability $q(x_0,n)$ for lattice random walk with $\pm 1$ 
jumps} 

{{ Here we consider a random walker on a $1$-d lattice (with integer points),
starting at the initial position $x_0$ (note that $x_0$ is a non-negative
integer). At each discrete time step, the walker jumps by either $+1$
or $-1$, chosen with equal probability $1/2$. This corresponds to a jump
PDF of the form, $f(\eta)= \frac{1}{2}\delta (\eta-1)+ \frac{1}{2} \delta(\eta+1)$, which 
is symmetric but not {\it continuous}. Hence, the general results derived in this
paper for symmetric and continuous jump density cannot be directly applied to this case.
However, this case can be worked out separately as we show in this Appendix. }}

{{
Let $q(x_0,n)$ denote
the survival probability, i.e., the probability that the walker, starting
at $x_0$ ($x_0\ge 0$), stays non-negative up to step $n$. One can  
easily write down
a recursion relation for $q(x_0,n)$
\begin{equation}
q(x_0,n)= \frac{1}{2}\left[q(x_0-1,n-1)+q(x_0+1,n-1)\right] \;, \quad x_0\ge 0
\label{lrw.1}
\end{equation}
with the boundary condition: $q(-1,n)=0$. The initial condition is:
$q(x_0,0)=1$ for all $x_0\ge 0$. To solve this recursion, we define the  
generating function
\begin{equation} 
{\tilde q}(x_0,s)= \sum_{n=0}^{\infty} q(x_0,n)\, s^n \;.
\label{lrw.2}   
\end{equation}
Then, the generating function satisfies the inhomogeneous equation for
all $x_0\ge 0$
\begin{equation}
{\tilde q}(x_0,s)= 1+ \frac{s}{2}\left[{\tilde q}(x_0+1,s)+{\tilde 
q}(x_0-1,s)\right]
\label{lrw.3}
\end{equation}   
with the boundary conditions: $q(x_0=-1,s)=0$ and $q(x_0\to \infty,s)$
non-divergent. By making a shift, ${\tilde q}(x_0,s)=
(1-s)^{-1}+{\tilde r}(x_0,s)$ that gets rid of the
inhomogeneous
constant term in Eq. (\ref{lrw.3}) and solving the resulting homogeneous
equation for ${\tilde r}(x_0,s)$, it is easy to see that the most general
solution to Eq. (\ref{lrw.3}) is given by
\begin{equation}  
{\tilde q}(x_0,s)= \frac{1}{1-s}+ A\, \left[\lambda_{+}(s)\right]^{x_0}
+ B\, \left[\lambda_{-}(s)\right]^{x_0}
\label{lrw.4}
\end{equation}
where $\lambda_{\pm}(s)= (1\pm \sqrt{1-s^2})/s$ and $A$ and $B$ are two
arbitrary constants. The boundary condition that ${\tilde q}(x_0\to
+\infty,s)$ is non-divergent forces the constant $A=0$.
The other
boundary condition ${\tilde q}(-1,s)=0$ fixes the constant
$B=-\lambda_{-}(s)/(1-s)$. Hence, the full exact solution is then given by
\begin{equation}  
{\tilde q}(x_0,s)=\frac{1-\left(\lambda_{-}(s)\right)^{x_0+1}}{1-s}=   
\frac{1}{1-s}\left[1-
\left(\frac{1-\sqrt{1-s^2}}{s}\right)^{x_0+1}\right] \;, \quad\, {\rm
for}\,\,
{\rm all}\quad x_0\ge 0\, .
\label{lrw.6}
\end{equation}
It is now easy to extract the asymptotic behavior of $q(x_0,n)$ for large 
$n$ by analyzing Eq. (\ref{lrw.6}) near $s=1$. We consider again the
two regimes (i) discrete `quantum' regime where $x_0\sim O(1)$
is fixed,
and $n\to \infty$  and (ii) `classical' scaling regime where $x_0\sim
\sqrt{n}$ as $n\to \infty$, separately. }}

\vskip 0.4cm 

{{ \noindent {\bf Regime I: $\bm{x_0\sim O(1)}$}. Setting $s=1-p$ and expanding
the right hand side of Eq. (\ref{lrw.6}) in powers of $p$ (while keeping  
$x_0$ fixed) gives
\begin{equation}
{\tilde q}(x_0,s=1-p) = \frac{\sqrt{2} (x_0+1)}{\sqrt{p}} - (x_0+1)^2 +
O(\sqrt{p})\, .
\label{lrw.7}
\end{equation}
Hence, inverting the generating function, one gets the leading asymptotic
behavior of $q(x_0,n)$ for large $n$ with fixed $x_0$ as
\begin{equation}
q(x_0,n)= \frac{1}{\sqrt{n}}\, U_{\rm lw}(x_0) \;, \quad {\rm where}\quad    
U_{\rm lw}(x_0)= \sqrt{\frac{2}{\pi}}\, (x_0+1)\,,
\label{lrw.8}   
\end{equation}
where the subscript '${\rm lw}$' stands for the lattice walk.
Thus, in the `quantum' regime, $q(x_0,n)$ behaves in the same way
as in Eq. (10) of the main paper valid for continuous and symmetric jump
distribution, except that the scaling function $U(x_0)$ is different and 
exactly given by Eq. (\ref{lrw.8}) for all $x_0\ge 0$. }}

\vskip 0.4cm

{{
\noindent {\bf Regime II: $\bm{x_0\sim \sqrt{n}}$ with $\bm{z= x_0/\sqrt{n}}$
fixed}. In this case, seeting $s=1-p$ with $p\to 0$, and $x_0\to \infty$
but keeping the product $x_0\sqrt{p}$ fixed, the right hand side of Eq.
(\ref{lrw.6}), to leading order, is given by
\begin{equation}
{\tilde q}(x_0,s=1-p)\approx \frac{1- e^{-\sqrt{2p} (x_0+1)}}{p}\, .
\label{lrw.9}
\end{equation}
Inverting the laplace transform with respect to $p$, one gets, to leading
order in the scaling regime
\begin{equation}
q(x_0,n) \to V_{\rm lw}\left(\frac{x_0}{\sqrt{n}}\right) \;, \quad {\rm
where}\quad V_{\rm lw}(z)= {\rm erf}\left(\frac{z}{\sqrt{2}}\right)\,. 
\label{lrw.10}
\end{equation}  
}}
{{To summarize, in the case of a $\pm 1$ lattice walk, one gets the following
large $n$ behaviors for the survival probability
\begin{equation}
q(x_0,n) \sim \left\{\begin{array}{rl}
\dfrac{1}{\sqrt{n}}\, U_{\rm lw}(x_0)   \, &\textrm{for}\,\,\quad
n\rightarrow +\infty\,\quad \textrm{and}\,\,\quad  x_0= O(1) \;,
    \\
\vspace{1.5mm}\\
V_{\rm lw}\left(\dfrac{x_0}{n^{1/2}}\right)
&\textrm{for}\,\,\quad n\rightarrow +\infty\,\quad \textrm{and}\,\,\quad
x_0=O(n^{1/2}) \;,
\end{array}\right.
\label{result.lrw_appB}
\end{equation}
where
\begin{eqnarray}
U_{\rm lw}(x_0) & =& \sqrt{\frac{2}{\pi}}\, (x_0+1) \quad {\rm for}\,\,
{\rm
all}\quad x_0\ge 0 \label{ux0_lrw} \;,\\
V_{\rm lw}(z) & =& {\rm erf}\left(\frac{z}{\sqrt{2}}\right) \;.
\label{v2z_lrw_appB}
\end{eqnarray}
}}
\vskip 0.4cm 

{{It is easy to check that the behaviors of $q(x_0,n)$ match
smoothly at the boundary between the two regimes I and II. Taking
$x_0 \gg 1$ in the discrete quantum regime, one gets $q(x_0,n)\approx
\sqrt{2/\pi}\, x_0$. On the other hand, letting
$x_0 \ll \sqrt{n}$ in the classical scaling regime and using
${\rm erf}(z)\to \frac{2}{\sqrt{\pi}}\, z$ as $z\to 0$, one gets
the same expression $q(x_0,n)\approx \sqrt{2/\pi}\, x_0$, ensuring
a smooth matching between the two regimes. {In Fig. \ref{Fig_crossover_lw}, we show a plot of $q(x_0,n)$ for a lattice random walk of $n=1000$ steps and compare it with our analytical predictions in Eqs. (\ref{ux0_lrw}) and (\ref{v2z_lrw_appB})}.} }

{}

\end{document}